\documentclass[aps,prb,twocolumn,superscriptaddress]{revtex4-1}

\include{ams}
\usepackage{amsmath,amssymb,amsthm}
\usepackage{dcolumn}
\usepackage{bm}
\usepackage{graphicx}
\usepackage{threeparttable}
\begin{document}

\title[Relaxation Time Spectrum of Low Energy Excitations in 1 and 2 D materials with Charge
- ore Spin - Density Waves]{Relaxation Time Spectrum of Low Energy Excitations in 1 and 2-D materials with Charge
- or Spin - Density Waves}

\author{S. Sahling}
\affiliation{Institut f\"{u}r Festk\"{o}rperphysik, TU Dresden,01062 Dresden, Germany}
\affiliation{CNRS,  Institute N\'{e}el, F-38042 Grenoble, France}
\author{G. Remenyi}
\affiliation{CNRS,  Institute N\'{e}el, F-38042 Grenoble, France}
\affiliation{Universit\'{e} Grenoble Alpes, Institute N\'{e}el, F-38042 Grenoble, France}
\author{J.E. Lorenzo}
\author{ P. Monceau}
\affiliation{CNRS,  Institute N\'{e}el, F-38042 Grenoble, France}

\author{V.L. Katkov}
\affiliation{Bogoliubov Laboratory of Theoretical Physics, Joint
Institute for Nuclear Research, 141980 Dubna, Moscow region,
Russia}

\author{V.A. Osipov}
\affiliation{Bogoliubov Laboratory of Theoretical Physics, Joint
Institute for Nuclear Research, 141980 Dubna, Moscow region,
Russia}

\begin{abstract}
The long-time thermal relaxation of (TMTTF)$_2$Br,  Sr$_{14}$Cu$_{24}$O$_{41}$ and Sr$_2$Ca$_{12}$Cu$_{24}$O$_{41}$ single crystals at temperatures below 1 K and magnetic field up to 10 T is investigated. The data allow us to determine the relaxation time spectrum of the low energy excitations caused by the charge-density wave (CDW) or spin-density wave (SDW). The relaxation time is mainly determined by a thermal activated process for all investigated materials.  The maximum relaxation time increases with increasing magnetic field. The distribution of barrier heights corresponds to one or two Gaussian functions. The doping of Sr$_{14-x}$Ca$_{x}$Cu$_{24}$O$_{41}$ with Ca leads to a drastic shift of the relaxation time spectrum to longer time. The maximum relaxation time changes from 50 s (x = 0) to 3000 s (x = 12) at 0.1 K and 10 T. The observed thermal relaxation at x=12 clearly indicates the formation of the SDW ground state at low temperatures.  
\end{abstract}

\pacs{}


\maketitle
\section{Introduction}

Low energy excitations exist in low dimensional systems (1D or 2D) with charge (CDW) or spin (SDW)
density waves  with a broad spectrum of relaxation time. This leads to a non-exponential thermal relaxation and an
additional contribution to the heat capacity below 1K which strongly depends on the time and magnetic field \cite{1c,2c,3c,4c,5c}. 

It was not possible to determine the equilibrium heat capacity of this contribution in
materials with incommensurate CDW or SDW up to now, since the maximum of the relaxation time spectrum was too long in
comparison to the time window of the experiment \cite{5c}. This was achieved for the first time with the commensurate 1D SDW
compound (TMTTF)$_2$Br by extending the time window of the measurement up to 22000 s \cite{6c,7c}. It was shown that the equilibrium heat capacity is strongly proportional to
$T^{-2}$ corresponding to an upper tail of a Schottky term with a magnetic field dependent Schottky energy $E_s$. The end of the inner relaxation
$t_{in}$ was obtained as a function of the temperature and magnetic field. The experimental parameter $t_{in}$ is
close to the maximum value of the relaxation time spectrum. It follows the Arrhenius law:
\begin{equation}\label{eq1}
t_{in} = \tau_0 \exp\left(\frac{E_a}{k_B T}\right),
\end{equation}
with an activation energy $E_a/k_B=0.5$ K that does not depend on the magnetic field $H$ while the parameter $\tau_0$ is found to be proportional to $H^2$ at large $H$. The maximum of the Schottky term always occurs at too low temperatures to be observed in this experiment. 

It should be noted that the long-time thermal measurements have proven highly effective in studies of the ground-state structure of a variety of materials at very low temperatures. In particular, a recent experiment~\cite{Pomar} reports the observation of the sharp heat capacity increase
in pyrochlore oxide Dy$_2$Ti$_2$O$_7$ below $0.4$ K that has not previously been detected. This led to a conclusion that the ground state of thermally equilibrated Dy$_2$Ti$_2$O$_7$ might correspond to the onset of order instead of the expected degenerate manifold of spin-ice states.

A new group of materials is Sr$_{14-x}$Ca$_{x}$Cu$_{24}$O$_{41}$ where a 2D CDW is formed after an insulator-to-insulator transition unlike the standard systems where electron density waves (DW) appear at temperatures below a metal-to-insulator transition. Increasing the Ca content destroys the order so that the CDW disappears completely for x$\geq$11.5. Notice that CDW dynamics in these materials has been intensively studied by using different experimental techniques including dc and electric field dependent transport, various spectroscopic methods, low-frequency electronic Raman scattering, and resonant X-ray diffraction (for details, see review~\cite{8c} and the references therein).    

Recently \cite{9c}, the heat capacity was investigated for $x = 0$ and $x = 12$. The equilibrium heat capacity and the parameter $t_{in}$ were obtained for both materials as a function of the temperature and magnetic field.  For the material without Ca, the Schottky contribution was found including the presence of the maximum of the Schottky term. A surprising result was obtained for $t_{in}$ which proved much shorter compared to  other investigated materials. In Ca-doped sample ($x = 12$), the Schottky term was also observed but  with significantly longer $t_{in}$. 
The maximum relaxation time of the spectrum was determined in all these experiments only. An interesting question is: how does the whole relaxation time distribution function look out? The aim of our paper is to answer this question.

\section{Experiment}

The relaxation time spectrum and the equilibrium heat capacity can be obtained by the relaxation time method. First, a constant heater power changes the sample temperature after the waiting time $t_w = 2 t_{in}$ to the new equilibrium temperature $T_1$. Then the heater power is switched off at $t = 0$ and the temperature is measured as a function of time. The equilibrium relaxation time $\tau_{eq}$ can be obtained for $t  > t_{in}$ (see Fig.~1) and we can determine the heat capacity $C_p = \tau_{eq}/R_{hl}$ with  $R_{hl} = (T_1-T_0)/(U_h I_h)$ being the thermal resistance between the sample and the sample holder, $U_h$ the heater voltage, and $I_h$ the heater current. 

The distribution function of the relaxation times $G(\ln\tau)$ can be recovered from the temperature dependence at short times $t < t_{in}$, which can be expressed as \cite{10c}
\begin{equation}\label{eq2}
\frac{\Delta T_i}{\Delta T_0} = \int\limits_{\ln \tau_{min}}^{\infty} G(\ln\tau)\exp\left(-\frac{t}{\tau}\right) d \ln\tau,
\end{equation}			
where $\tau_{min}$ is a short-time cutoff,
\begin{equation}
\Delta T_i(t) = \Delta T(t) - \Delta T_{eq},  
\end{equation}
and
\begin{equation}\label{eq4}
\Delta T_{eq} = \Delta T_{eq}^0 \exp\left(-\frac{t}{\tau_{eq}}\right).
\end{equation}
Namely, we subtract from the measured temperature difference at given time its extrapolated quasi-equilibrium value. Notice that this procedure brings additional numerical error which, together with the error in measurements, does not exceed 10 percent for $\Delta T_i(t)$. A similar estimate is valid for all extracted parameters.

We are looking for a distribution function of $\ln\tau$, which corresponds to a distribution of the barrier heights in the case of thermal activation or tunneling process. We will show that the data of all investigated materials can be calculated with one or two Gaussian distribution functions. We get reasonable results only if we measure the time dependence of the temperature relaxation longer than $t_{in}$, which means that  $\tau_{eq} = R_{hl} C_p$  must be longer than $t_{in}$. For this reason, we have to choice a very large heat link $R_{hl}$ and sample mass. On the other side, the maximum time of measurements is limited by the fluctuation of temperature caused by fluctuation of the parasitic heat flow:
\begin{equation}
\Delta T(t_{max})= \Delta T_{fl} = R_{hl}\Delta P_{par}.
\end{equation}

The experiment required a very low value of the parasitic heat flow fluctuations during a long time.
$\Delta P_{par}$ was in our experiments between 1 and 10 pW. This leads to a temperature fluctuation in a range  between 10 and 100 $\mu$K, where an optimal heat leak of $10^7$ K/W at 0.1 K was used.  To realize the heat link the sample was mounted together with a heater and a self-made AuGe-thermometer \cite{11c} on the end of a thin Si platelet (30 $\times$ 4 $\times$ 0.2 mm$^3$), which gives a heat link of $10^8$ at 0.1 K. The heat link is than reduced by a 30 $\mu$m Cu wire on the thermometer. The sample was in the form of a sandwich: heater-sample-thermometer-heat link. All together was fixed by small Teflon threads and grease. The addenda was measured separately. The schematic picture of the device is shown in Fig. 1. 
\begin{figure}
\includegraphics[width=6cm]{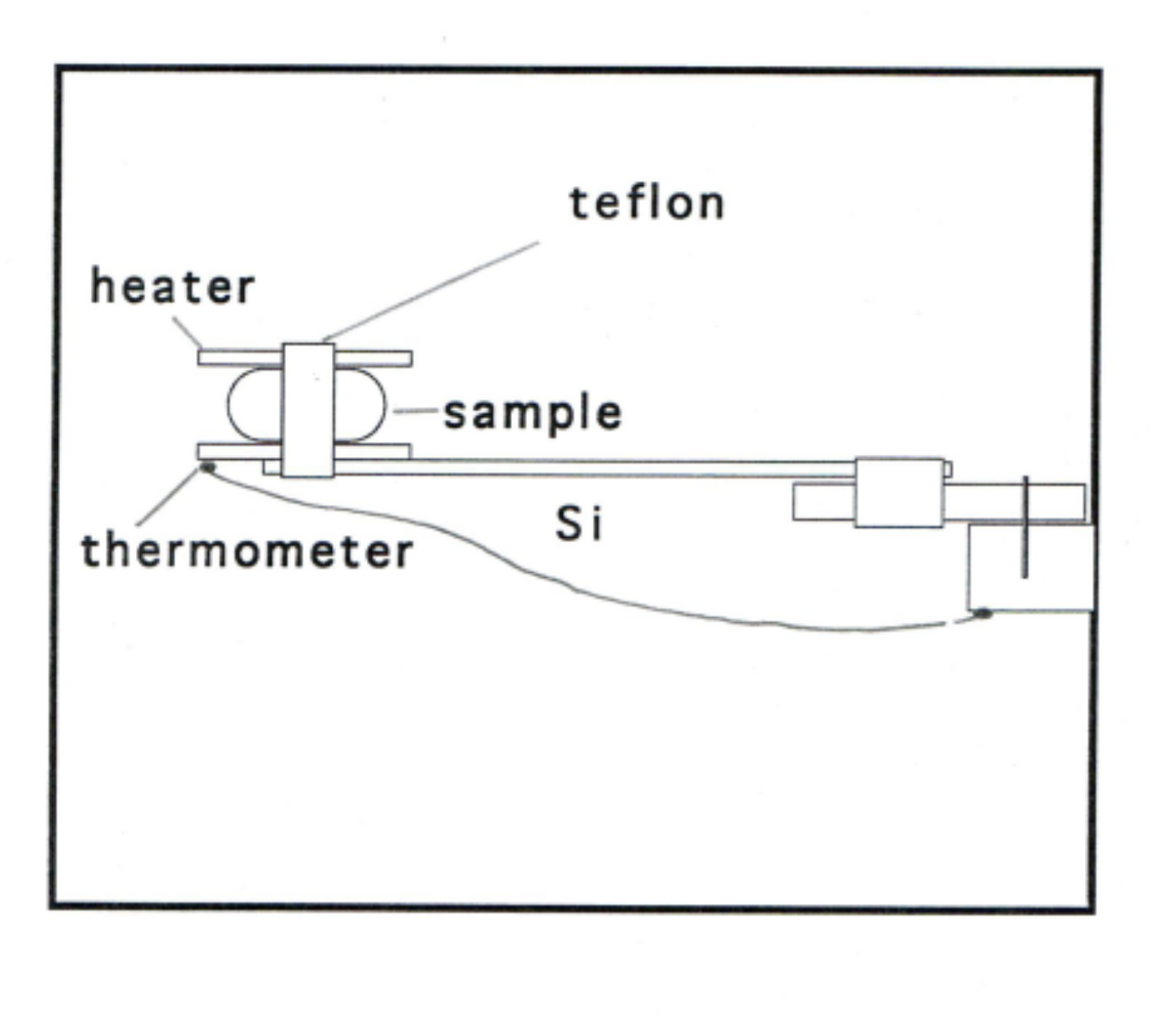} 
\caption{Schematic picture of the experimental setup.}
\end{figure}
In all experiments single crystals were used with masses 64 mg (TMTTF)$_2$Br, 440 mg (Sr$_{14}$Cu$_{24}$O$_{41}$), and 2082 mg (Sr$_2$Ca$_{12}$Cu$_{24}$O$_{41}$).

\section{Results}
\subsection{(TMTTF)$_2$Br}

Fig.~2 shows the thermal relaxation from $T_1 = 102$ mK to $T_0 = 92$ mK after a constant power of the sample heater is switched off at $t = 0$ at different magnetic field.  
\begin{figure}
\includegraphics[width=8cm]{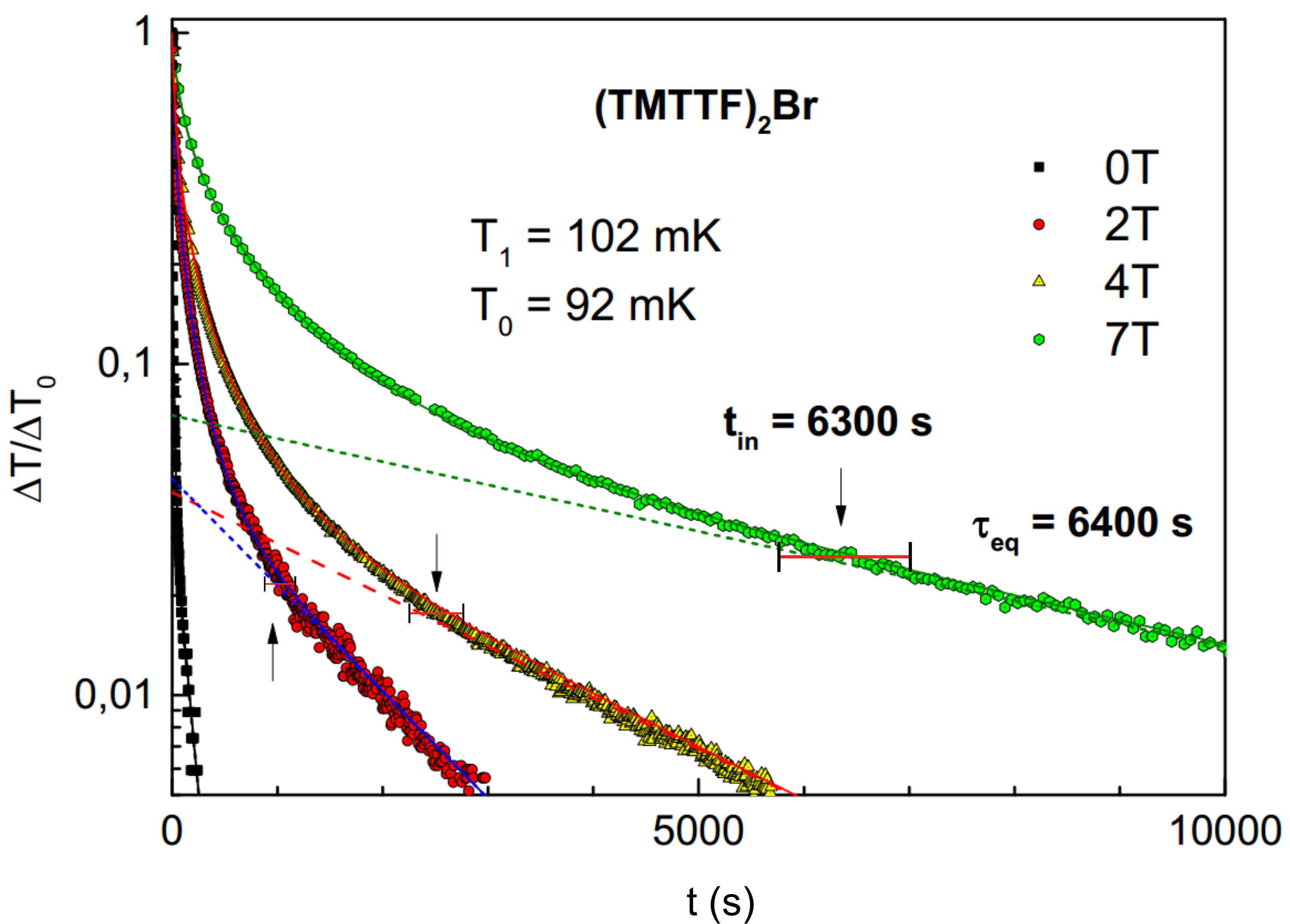} 
\caption{A constant power of the sample heater is switched off at $t = 0$ after the waiting time $t_w = 2 t_{in}$ and the temperature relaxes from $T_1 = 102$ mK to $T_0 = 92$ mK. The inner relaxation is finished after the time $t_{in}$ (arrows). 
The thermal relaxation follows Eq. (\ref{eq4})  at longer time $t > t_{in}$ (straight lines). The quasi-equilibrium relaxation time $\tau_{eq}$ and the end of the inner relaxation $t_{in}$ depend strongly on the magnetic field.}
\end{figure}
The magnetic field was directed along the c-axis (parallel to the SDW). The waiting  time of the constant power was $2 t_{in}$, so that we start from the equilibrium state at temperature $T_1$. The temperature difference $\Delta T_0 = T_1-T_0$ was between 5 and 10 \% of $T_0$. The non-exponential dependence of the time at the beginning is caused by the inner relaxation process, which is finished at $t_{in}$ (arrows in Fig.~2). At longer time the thermal relaxation follows Eq. (\ref{eq4}) (strait lines) and the parameters  $\Delta T_{0eq}$ and $\tau_{eq}$ can be determined. The relaxation time  $\tau_{eq}$  yields together with the corresponding heat link the equilibrium heat capacity, which is in good agreement with our earlier results \cite{6c,7c}.  Another presentation of the thermal relaxation is given in Fig.~3.  
\begin{figure} \label{fig2}
\includegraphics[width=8cm]{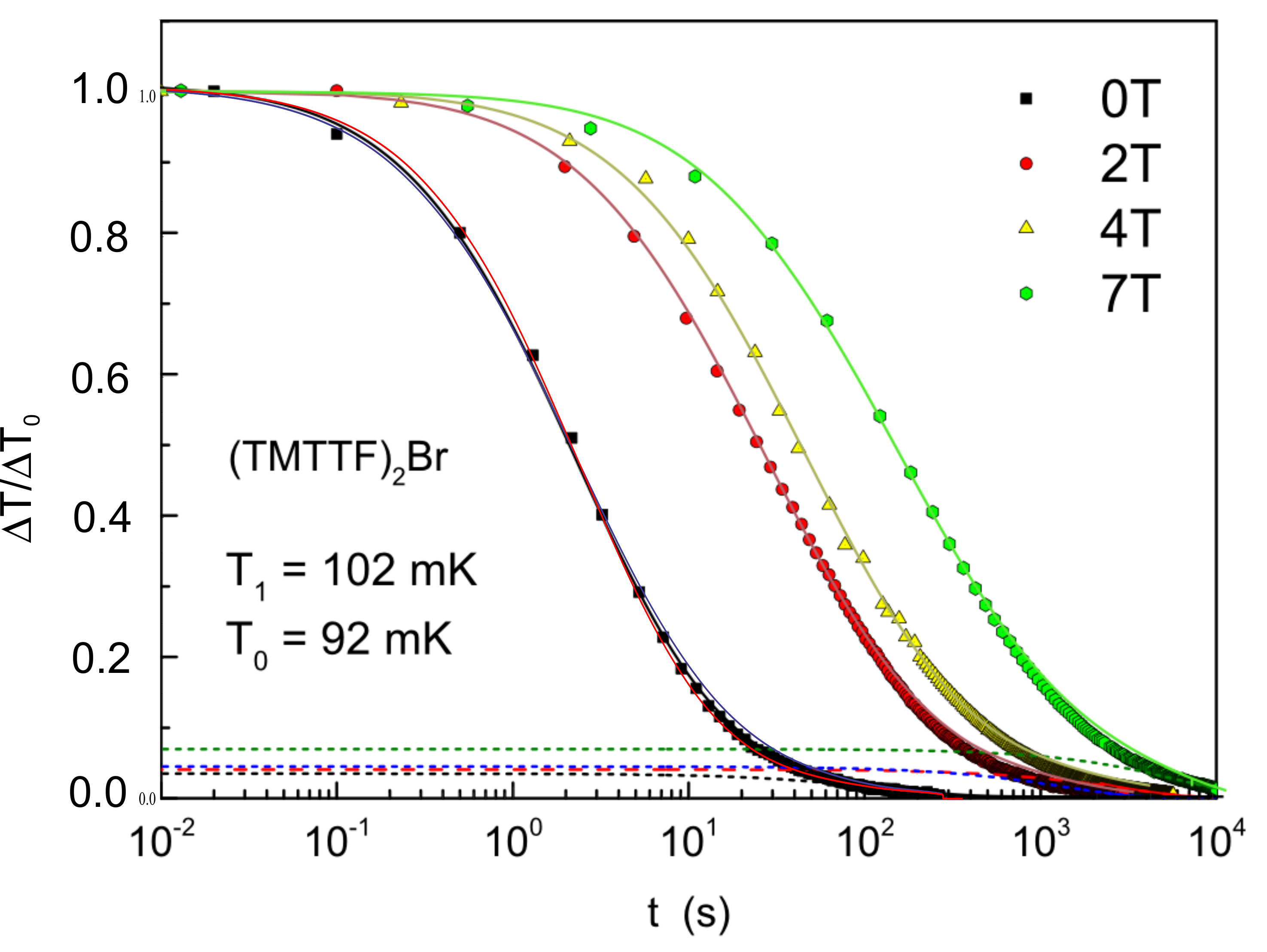}
\caption{The data of Fig.~2 in the semi-log presentation. The straight lines of Fig.~2 (quasi-equilibrium relaxation) are shown by the dashed curves. More than 90\% of the measured initial temperature difference is caused by the inner relaxation process. The solid lines are calculated by Eqs.~(\ref{eq2})-(\ref{eq4}) with a Gaussian distribution function. Two fitting parameters of this distribution function $\tau_m$ (the position of the maximum) and $w$ (the width of the distribution) are given in Table I and Fig. 5 for $T_0 = 0.092$ K. For zero field, two additional fitting curves show the cases of $1.1 w$ (blue) and $0.9 w$ (red).}
\end{figure}
It is seen that the maximum value of $\Delta T_{0eq}$ is about 5-7 \% of the total  $\Delta T_{0}$. This means that we have less than $0.7$ mK to determine the exponential relaxation law and the equilibrium heat capacity at $0.1$ K.

The process of relaxation is generally characterized by the relaxation rate
\begin{equation}
S(t) =  \frac{d}{d \ln t}\left(\frac{\Delta T_i}{\Delta T_{0}}\right).
\end{equation}
Differentiation of the right side in Eq. (\ref{eq2}) with respect to $\ln t $  converts the kernel of integration $\exp(-t /\tau)$  to a peak $-\exp(-e^{\ln t - \ln \tau} + \ln t - \ln \tau)$  with a center at $\ln t $, the height $1/e$ and the width of approximately $e$ in the scale of $\ln \tau$. In the case where $G(\ln \tau) $ changes slowly, the peak may be considered as a $\delta$-function, so that $G(\ln t) = S(\ln t)$. It is quite appropriate to describe data for (TMTTF)$_2$Br. However, if $G (\ln t)$ is a rapidly varying function, $S(t)$ gives the broadened distribution  of relaxation times. 
As will be shown, this occurs in the other two materials where the distribution functions look like sharp peaks. For this reason, in order to obtain $G(\ln \tau)$ properly we use another procedure. Namely,
we assume that $G(\ln \tau)$ is given by a Gaussian distribution
\begin{equation}\label{eq7}
G_i(\ln\tau) = \frac{A_i}{w_i\sqrt{2\pi}} \exp\left(-\frac{\ln^2(\tau/\tau_{mi})}{2 w_i^2} \right),
\end{equation}
where $A_i = \Delta T_i/\Delta T_{0}$ at $t = 0$. Then we replace the integral in  Eq.~(\ref{eq2}) by  a finite sum. As a result, one gets a polynomial $R(t,\tau_{mi}, w_i)$ with $t$ being a variable and $\tau_{mi}$ and $w_i$ the fitting parameters. These parameters are determined to provide a best fit of $R(t,\tau_{mi}, w_i)$ to the experimental data for $\Delta T_i(t)/\Delta T_{0}$  in Fig.~3. The result is given in Table I. As a proof, we substitute the Eq.~(\ref{eq7}) with extracted $\tau_{mi}$ and $w_i$ to the right-hand side of Eq.~(\ref{eq2}). The result of numerical integration is shown in Fig.~3 by solid lines. Notice that the relative error is found to be less than 3 percent in all considered cases. 
\begin{figure}
\includegraphics[width=8cm]{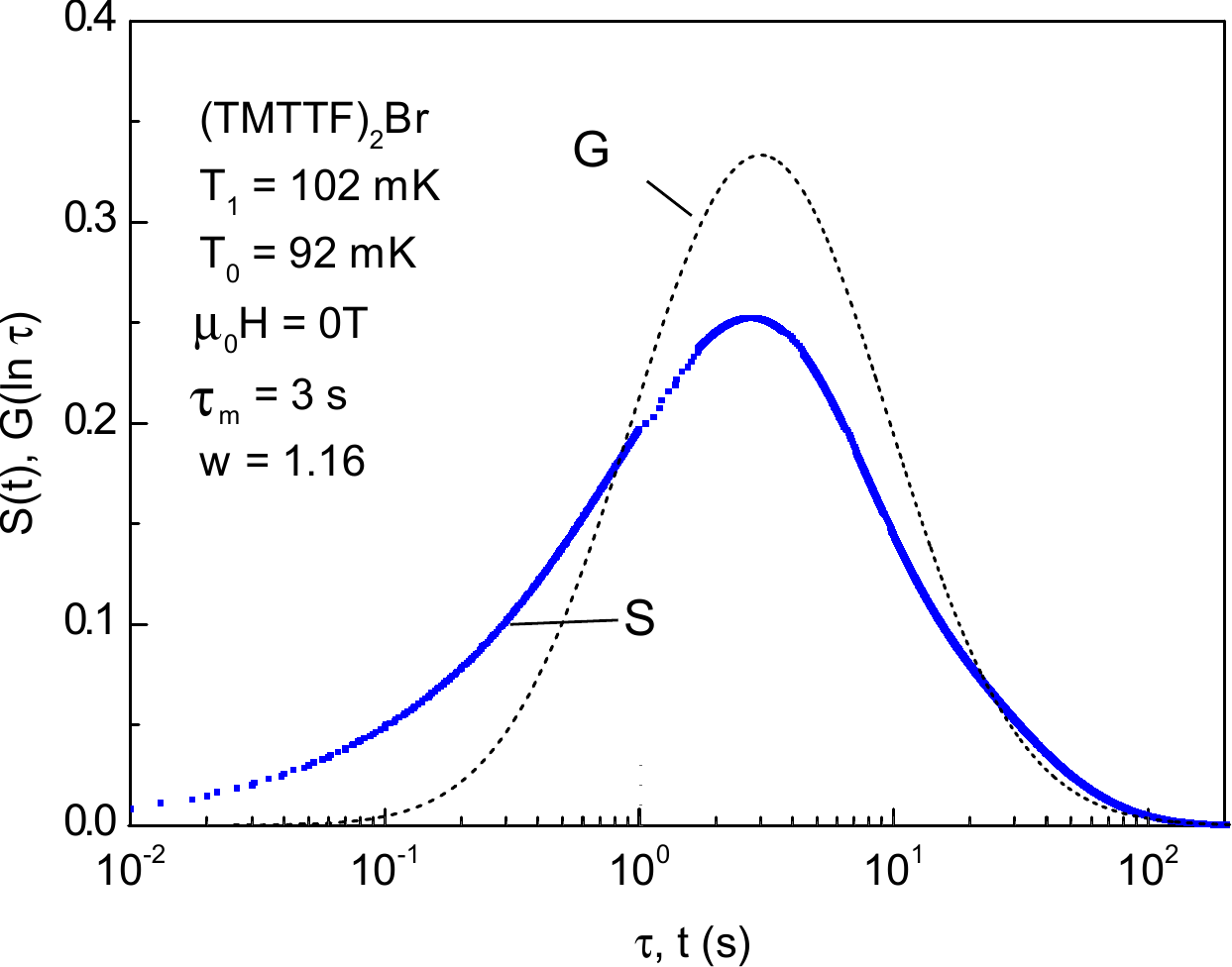}
\caption{The measured inner thermal relaxation $\Delta T_i/\Delta T_{i0}$ in the presentation $d(\Delta T_i/\Delta T_{i0})/d(\ln t)$ as a function of time (blue points) for the data shown in Figs.~2 and 3.  This gives roughly the distribution function of relaxation time (see the text). The numerical calculation of $\Delta T_i/\Delta T_{i0}$ with Eq.~(\ref{eq2}) yields a Gaussian distribution (dashed curve) with nearly the same position of the maximum $\tau_m$. 
}
\end{figure}
As is seen in Fig.~4, the measured relaxation rate becomes slightly broadened as compared to calculated $G(\ln \tau)$ while the position of the maximum is almost identical. To illustrate the sensitivity of the fitting procedure to the parameter $w_i$, the additional curves with a 10 percent shift of the width of the distribution are show in Fig.~3 for zero field: $0.9 w_i$ (red line) and $1.1 w_i$ (blue line). 
Similar deviations are found for non-zero fields.  

\begin{table}\label{tb1}
\caption{Parameters of the Gaussian distribution  
	for (TMTTF)$_2$Br. Estimated errors for the given
	values are $10$~\%. 
	}
\begin{tabular}{|c||r|r|r|r|}
\hline
$\mu_0 H$& ~~~0 T~~~ & ~~~2 T~~~ & ~~~4 T~~~ & ~~~7 T~~~ \\
\hline
\hline
$A_i$ & 0.965 &0.955 & 0.959 & 0.93 \\
\hline
$w_i$ & 1.16 & 1.35 & 1.37 & 1.46 \\
\hline
$\tau_{mi}~[s]$ & 3.0 & 36 & 64.9 & 196 \\
\hline
\end{tabular}
\end{table}

Parameters $t_{in}$ and $\tau_{mi}$ have the same magnetic field dependence
\begin{eqnarray} \label{x2}
t_{in}   & = & t_{in}(0)+ B_i(\mu_0 H)^2,\\
\tau_{mi}& = & \tau_{mi}(0) + C_i(\mu_0 H)^2,\label{x3}
\end{eqnarray}	
where $t_{in}(0)= 2.5\alpha$ s, $B_{i} = 0.46\alpha$ sT$^{-2}$, $\tau_{mi}(0) =0.05\alpha$ s, $C_i = 0.015\alpha$ sT$^{-2}$, and $\alpha = \exp[0.51(\textrm{K})/T]$. The values are given for the field induced phase (see Ref.~\cite{7c}).
Meanwhile, the width of the distribution increases weakly linearly (see Fig.~5).  

\begin{figure}
\includegraphics[width=8cm]{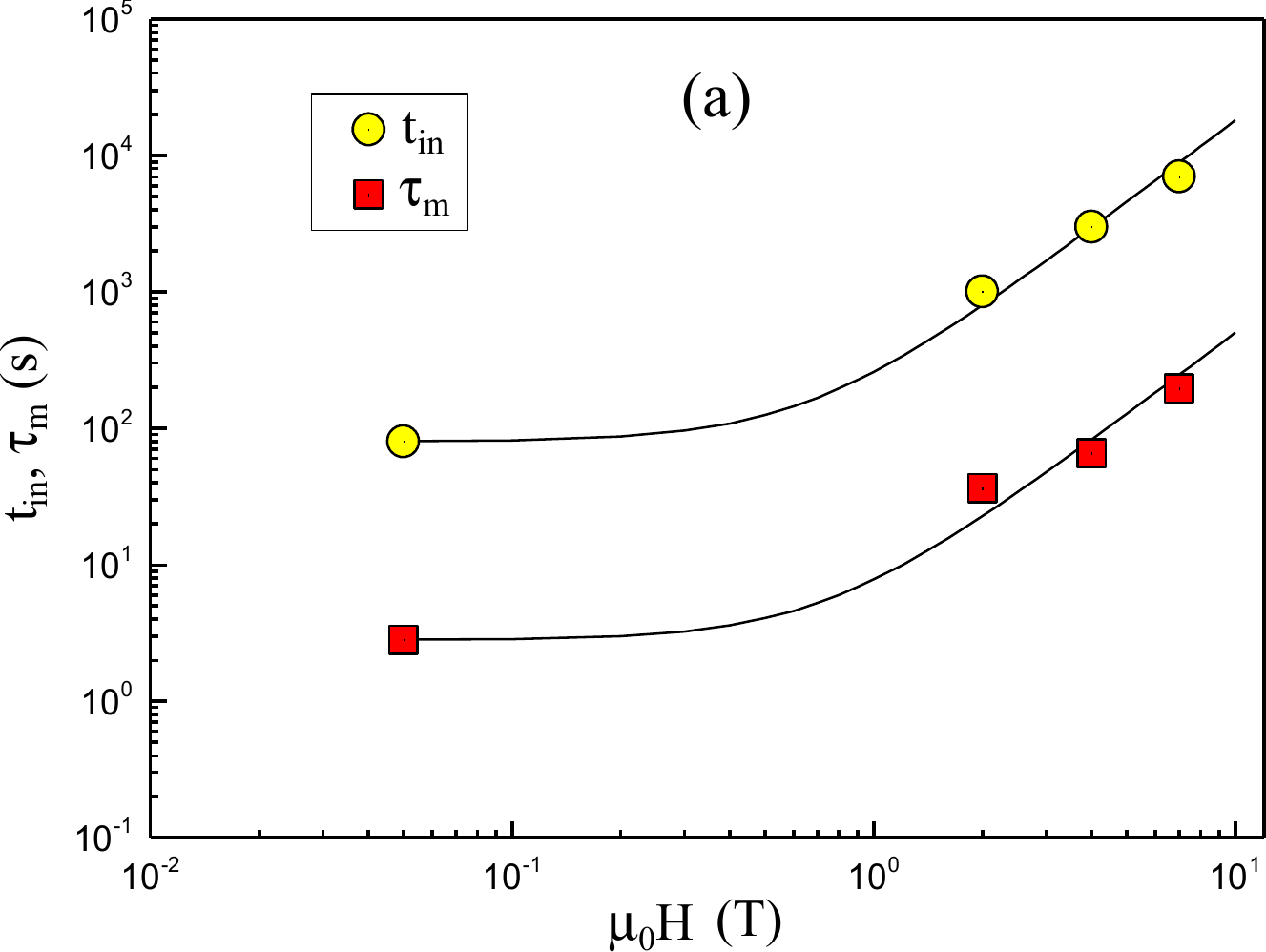}
\includegraphics[width=8cm]{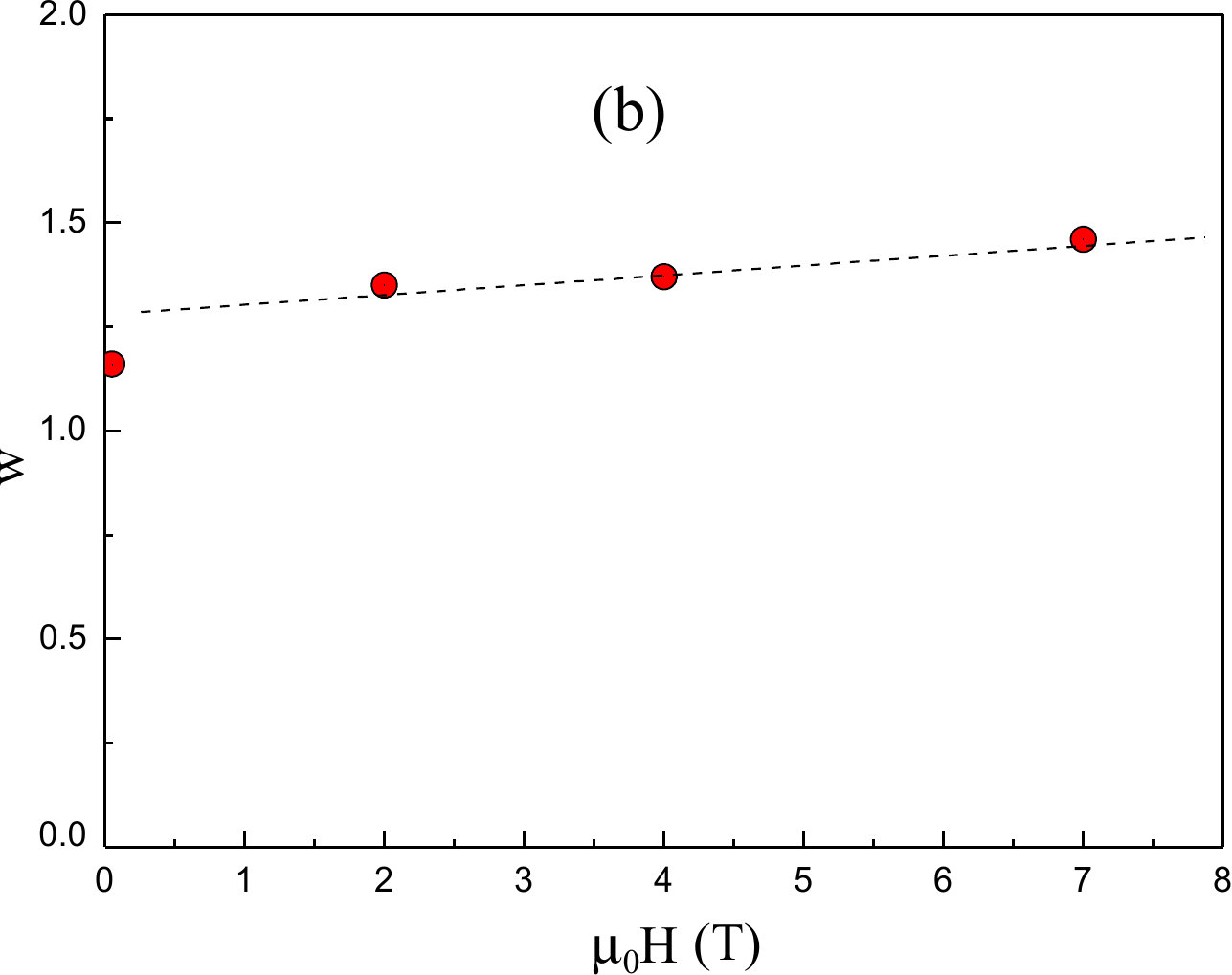}
\caption{
Eq.~(\ref{eq2}) yields a good fit to the inner thermal relaxation $\Delta T_i(t)$ when a Gaussian distribution function of relaxation time is used. The obtained fitting parameters $\tau_m$ (the position of the maximum) and $w$ (the width of the distribution) are shown in the Fig. 5a and b for the data of Fig. 3 ($T_0 = 0.092$ K and different mangetic field). Fig.~5a also shows the time $t_{in}$ (for $t > t_{in}$ the inner relaxation can be neglected, see Figs. 2 and 3). $\tau_m$ and $t_{in}$ are proportional to  $H^2$ at high magnetic field, i.e. with increasing magnetic field the distribution function is shifted  to higher values without change of the distribution width. The error corresponds to the size of symbols.}
\end{figure}

We have also determined the spectrum for different temperatures and fixed magnetic field. The result is the same as for $t_{in}$ in Ref. \cite{7c}: maximum values follow the Arrhenius law with the activation energy $E_a/k_B = 0.51$~K. $E_a$ does not depend on the magnetic field.

\subsection{Sr$_{14}$Cu$_{24}$O$_{41}$}

The heat capacity was measured by pulse technique in both   a very short ($30$ ms) and long time range with $\tau_{eq} > t_{in}$. The magnetic field was directed along the c-axis. 
\begin{figure}
	\includegraphics[width=8cm]{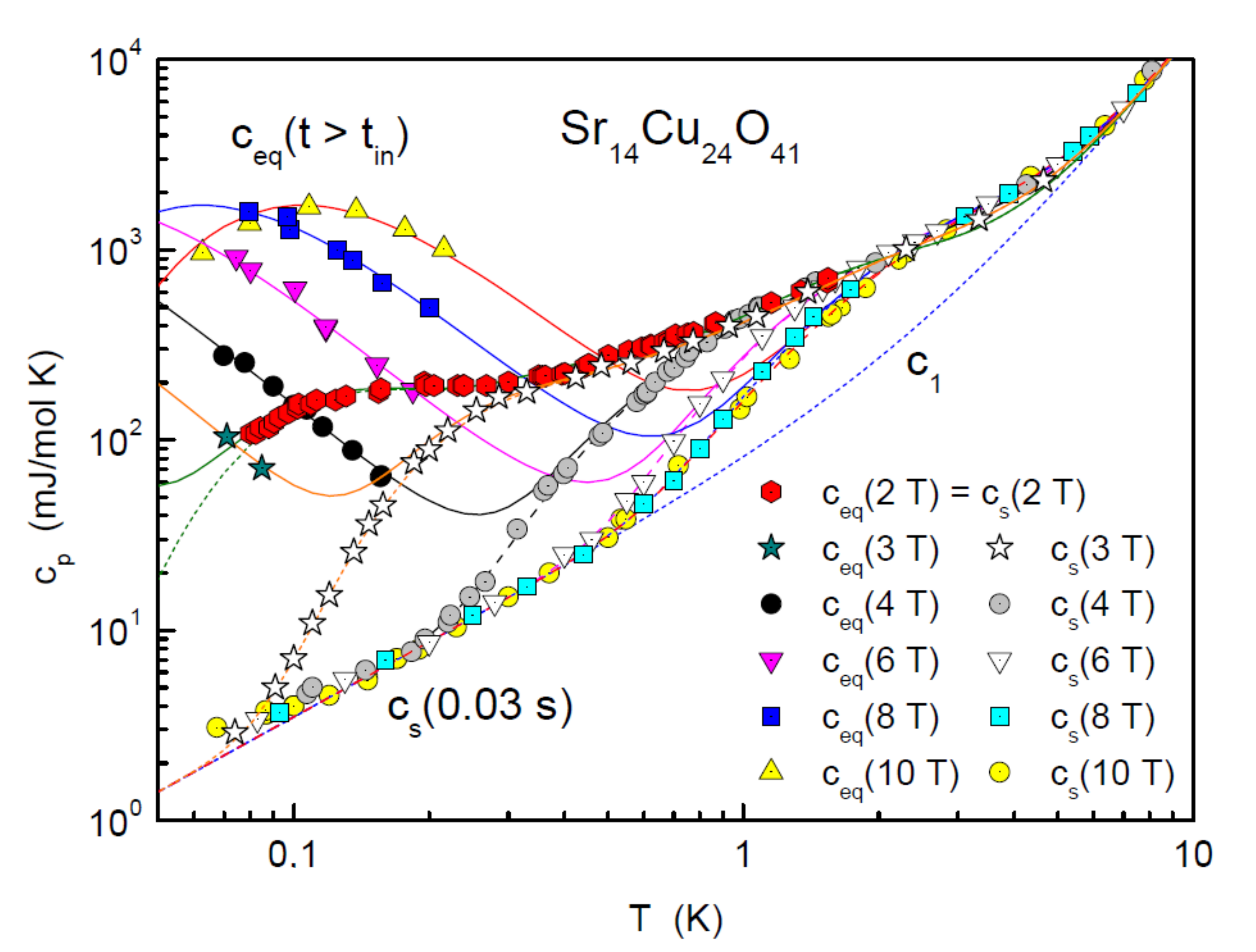}
	\caption{The heat capacity of Sr$_{14}$Cu$_{24}$O$_{41}$ as a function of the temperature for different magnetic fields measured at very short time ($t = 30$ ms) and long time ($t > t_{in}$). The CDW does not contribute to the short time heat capacity.  The solid lines show the total equilibrium heat capacity at different magnetic field. Below $0.3$ K it is determined by a magnetic field independent quasilinear term (dashed line) and a Schottky contribution. The Schottky energy is proportional to $H^2$.}
\end{figure}
The upper curves in Fig.~6 present the total equilibrium heat capacity including the Schottky term caused by the CDW. The contribution of the CDW can be completely excluded in the short time experiment thus allowing us to determine the background heat capacity (see Fig.~6). In contrast to (TMTTF)$_2$Br, there are many other (fast) contributions including 1D bosons \cite{12c}, which will be presented and discussed in a separate paper \cite{9c}. The CDW contribution is small in comparison to the background below $3$ T.  The maximum of the Schottky term was first observed in a magnetic field of $10$ T. The amplitude $A_s$ is equal to $3.9$ J/mol K. This value is essentially smaller in comparison to (TMTTF)$_2$Br where the estimated upper limit is about $50$ J/mol K. The Schottky energy is equal to $250$ mK at $10$ T and varies with the magnetic field as $H^2$.

The most surprising result is associated with $t_{in}$, which was determined in good agreement with the pulse and relaxation time method (see Fig. 7). 
\begin{figure}
	\includegraphics[width=8cm]{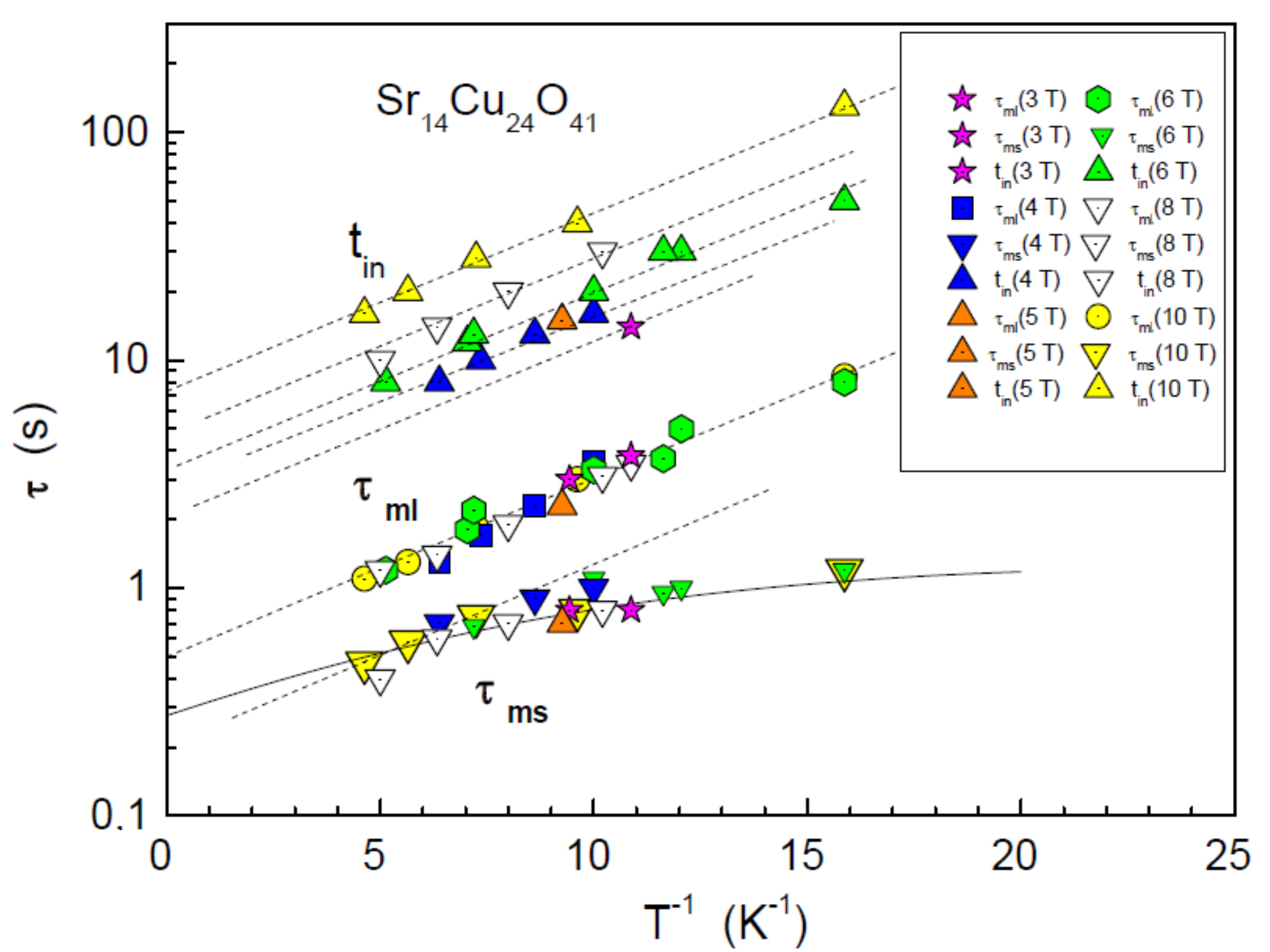}
	\caption{The end of the inner thermal relaxation $t_{in}$ (see also Fig.~8) follows the Arrhenius law (Eq.~\ref{eq1}) with a magnetic field independent activation energy $E_a/k_B = 0.2$ K. The absolute value depends weakly on the magnetic field. Two Gaussian distribution functions of relaxation time are necessary to calculate the inner relaxation $\Delta T_i(t)$ (according to Eq.~\ref{eq2}) with the maximum at $\tau_{ms}$ and $\tau_{ml}$ (see also Fig.~9). Both parameters are magnetic field independent, i.e. the increase of $t_{in}$ at fixed temperature and increasing magnetic field is a consequence of an increasing width. The temperature dependence of $\tau_{ml}$ corresponds to Eq.~(\ref{eq1}) with the same activation energy as for $t_{in}$. The temperature dependence of the shorter $\tau_{ms}$ deviates from Eq.~(\ref{eq1}) at low temperatures. The solid line is obtained when we include in addition a tunneling process (see Eqs.~(\ref{eq9}-\ref{eq10-5})).
	}
\end{figure}
The inner relaxation is finished at $0.1$ K and $10$ T even after $40$ s while the expected value for (TMTTF)$_2$Br  is $20000$ s ($10000$ s at $7$ T)! The temperature dependence again meets the Arrhenius law with a smaller activation energy $E_a/k_B = 0.2$ K. The magnetic field dependence is less strong: the value of $t_{in}$ is changed 3 times only with increasing magnetic field from $4$ T to $10$ T.

Fig. 8 shows the thermal relaxation after switching off a constant power of the sample heater ($T_1 = 112$ mK, $T_0 = 104$ mK, $\mu_0 H = 10$ T). 
\begin{figure}
	\includegraphics[width=8cm]{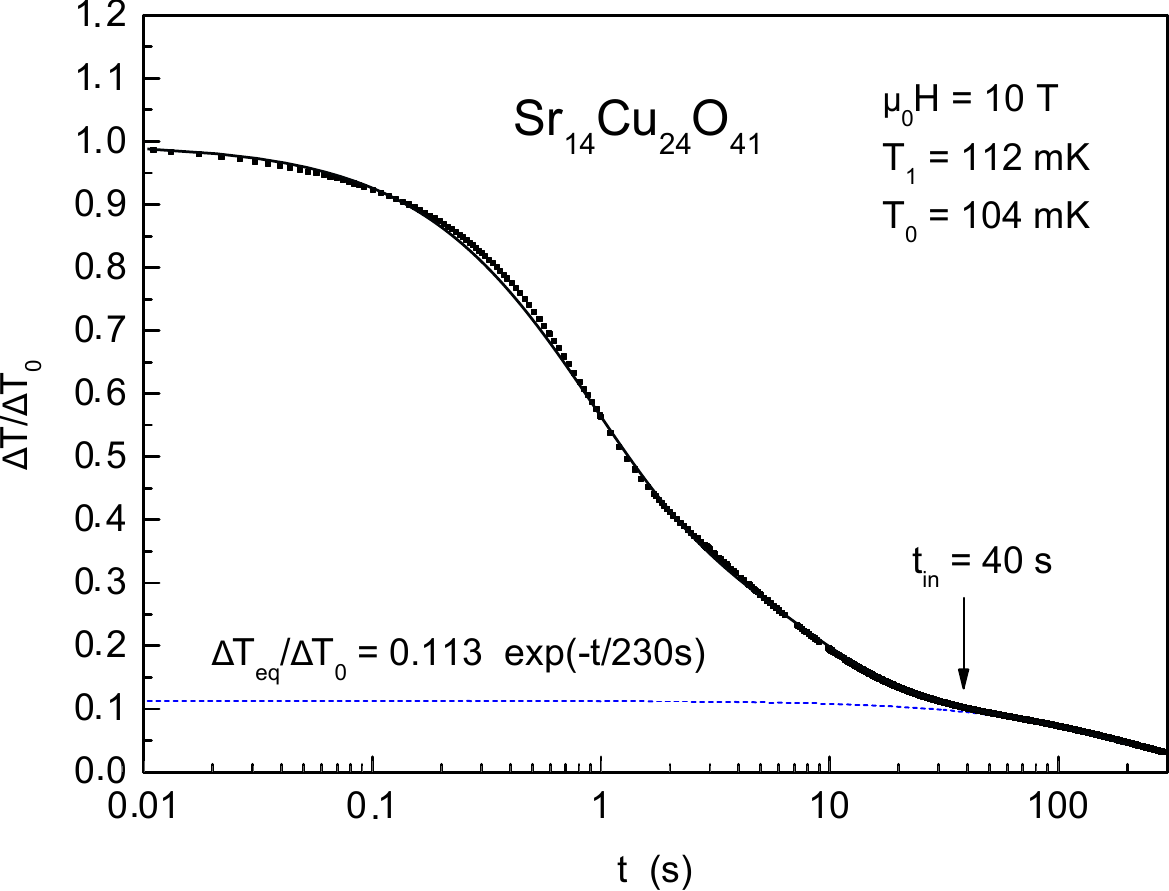}
	\caption{A constant power of the sample heater is switched off at $t = 0$ after the waiting time $t_w = 2 t_{in}$ and the temperature relaxes from $T_1 = 112$ mK to $T_0 = 104$ mK. The inner relaxation is finished after the time $t_{in}$ (the arrow). 
		The thermal relaxation follows Eq.~(\ref{eq4})  at longer time $t > t_{in}$ (dashed curve). The solid line shows the result of the calculation $(\Delta T_{eq} + \Delta T_i)/\Delta T_0$ where $\Delta T_i$ is obtained from Eq. (\ref{eq2}) with two Gaussian functions in Fig.~9 (dashed lines).}
\end{figure}
The exponential relaxation is observed for $t > 40$ s. The extrapolation to $t = 0$ gives a maximum of $\Delta T_{eq}$ which is estimated as $12$ \% of the total temperature difference $\Delta T_0$. Thus, for the determination of $\tau_{eq}$ and the equilibrium heat capacity remains less than $1$ mK. The other $88$ \% of $\Delta T$ is caused by the inner relaxation process (see Fig.~7). The same procedure as described above gives us the rough relaxation time spectrum (points in Fig.~9) and the best approximation of the relaxation time spectrum in the form of two Gaussian functions (dashed curves in Fig.~9). 
\begin{figure}
	\includegraphics[width=8cm]{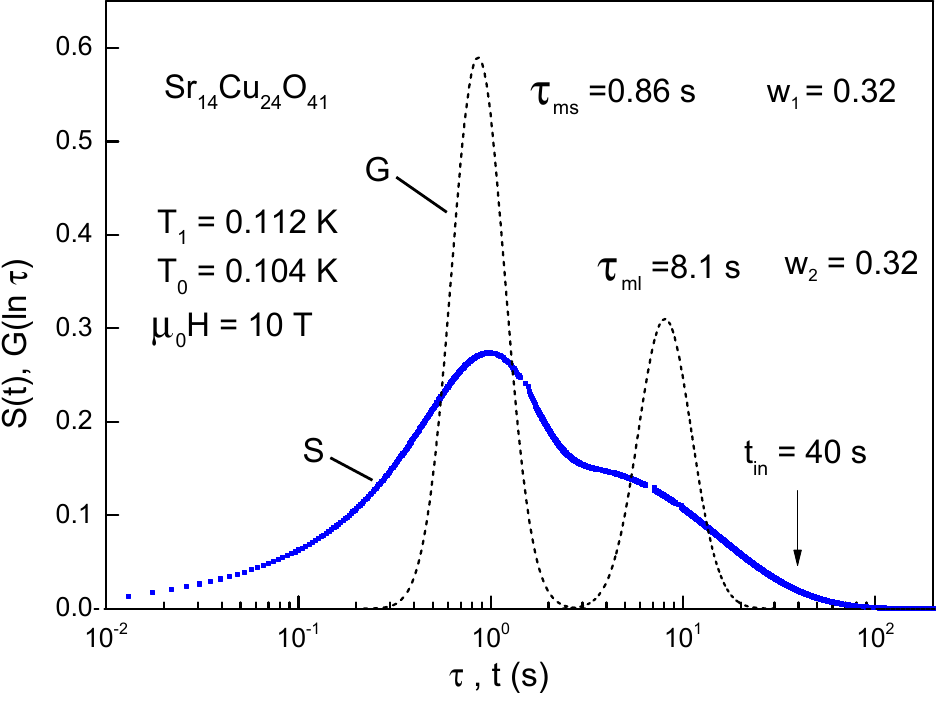}
	\caption{
		The measured inner thermal relaxation $\Delta T_i/\Delta T_{i0}$ in the presentation $d(\Delta T_i/\Delta T_{i0})/d(\ln t)$ as a function of time (blue points) for the data shown in Fig.~8.  This gives roughly the distribution function of relaxation time (see the text). The numerical calculation of $\Delta T_i/\Delta T_{i0}$ with Eq.~(\ref{eq2}) yields two Gaussian distribution (dashed curve). The corresponding four fitted parameters are shown. The positions of the peaks do not depend on the magnetic field (see Fig.~7).
	}
\end{figure}
The investigation of the magnetic field dependence at fixed temperature yields the next surprise: the positions of two distribution peaks do not change in the magnetic field. The upper distribution function becomes only broader and is responsible for a magnetic field dependence of $t_{in}$. The lower Gaussian function becomes unchanged in the magnetic field (see Fig.~7). In addition, the temperature dependence of $\tau_{ms}$ is nearly constant below $0.1$ K thus indicating that some tunneling process starts instead of the thermal activation.

\subsection{Sr$_{2}$Ca$_{12}$Cu$_{24}$O$_{41}$}

The heat capacity is shown in Fig.~10 as a function of the temperature for different magnetic field strengths. 
\begin{figure}
	\includegraphics[width=8cm]{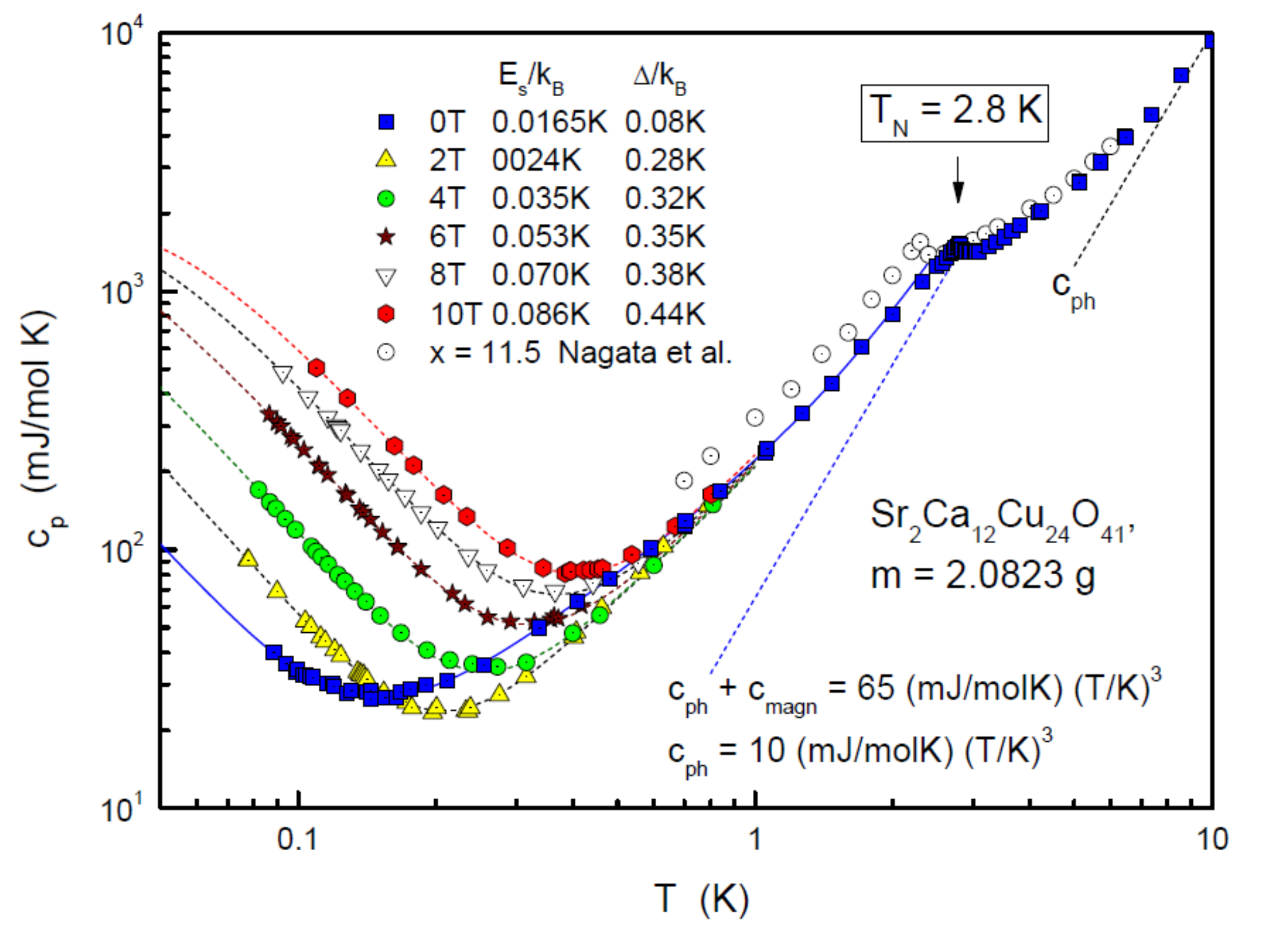}
	\caption{The heat capacity of Sr$_2$Ca$_{12}$Cu$_{24}$O$_{41}$ measured at long time $t > t_{in}$ as a function of the temperature for different magnetic field. Antiferromagnetic ordering is observed at $2.8$ K. The heat capacity is determined below the transition by a $T^3$ term (magnons and phonons), quasi-linear term with a magnetic field dependent gap, and the Schottky term which is strongly time dependent for $t < t_{in}$. The values of the gap $\Delta$ and Schottky energy $E_s$ for different magnetic field are given in the inset. The solid lines show the calculated equilibrium heat capacity, which is determined below $0.3$ K by the quasilinear term and the Schottky contribution only.}
\end{figure}
The data are obtained by the pulse technique. The magnetic field was directed perpendicular to the c-axis. A magnetic ordering is observed at $2.8$ K. The heat capacity is determined below this temperature by a $T^3$ term (phonons and magnons), a quasilinear term, and a Schottky term. A gap opening is observed for the quasilinear term between $0$ and $2$ T. In particular, this leads to a crossing of the curves for $0$ and $2$ T (see Fig.~10). Gap values are also given in Fig.~10 (for more details, see Ref.~\cite{9c}).

As in the previous case, the Schottky term is time dependent. Fig.~10 shows the equilibrium heat capacity only. The condition $\tau_{eq} > t_{in}$ is satisfied for all the experimental values of temperature and magnetic field due to the high background heat capacity and the large sample mass. We fit the Schottky term with the amplitude of the Sr$_{14}$Cu$_{24}$O$_{41}$ sample since the maximum is at lower temperatures. We get a good fit to the data (see solid lines in Fig.~10). The Schottky energy is the last free parameter and it is also given in Fig.~10. The Schottky energy depends only weakly on the magnetic field (a roughly linear dependence) in difference to other two materials.

The parameters of thermal relaxation at $0.1$ K and different magnetic field strengths are given in Table~II, where $\Delta T_{eq}$ is subtracted. At high temperature, this contribution is relatively large that does not yield the accurate amplitude of the distribution function. However, this can be easily corrected as we have found that the distribution function is determined by one Gaussian only with both magnetic field and temperature independent width $w$ (see Fig.~11c where dots show $d(\Delta T_i/\Delta T_{i0})/d(\ln t)$ and dashed lines show the Gaussian function). Similarly we obtain the distribution function at $10$ T and different temperatures (see Fig.~12). The width of the distributions remains always constant ($w=0.5$). Thus the maximum of the distribution shifts with magnetic field and/or temperature. This parameter is given in Fig.~13 as a function of $1/T$ for different magnetic field strengths. In contrast to other two materials, the activation energy is not a constant but increases with increasing magnetic field. Deviations from the Arrhenius law are found at low temperatures. The curves are calculated using Eq.~(\ref{eq9})
\begin{equation}\label{eq9}
\tau^{-1} = \tau^{-1}_{ta} + \tau^{-1}_{tun},
\end{equation}
with the relaxation time of the thermal activated process
\begin{equation}\label{eq10}
\tau_{ta}= \tau_{0} \exp\left[\frac{E_a(H)}{k_B T} \right], 
\end{equation}
and the relaxation time of tunnelling
\begin{equation}\label{eq10-5}
\tau_{tun}  = \frac{\tau_{tun}(0.1 \textrm{K})}{(T/0.1 \textrm{K})^3},
\end{equation}
$\tau_{tun}(0.1K) = 70 \exp(0.21 \mu_0 H)$ s, where $\mu_0H$ is in Tesla units. 
Both fitting parameters, $E_a$ and $\ln [\tau_{tun} (0.1 \textrm{K})]$, are proportional to the magnetic field. This is not surprising since both values are proportional to the barrier height. We get for the activation energy
\begin{equation}
E_a/k_B = [0.79 + 0.017 (\mu_0H)]~\textrm{K},
\end{equation}
where $\mu_0H$ is in Tesla units.
\begin{figure}
	\includegraphics[width=7 cm]{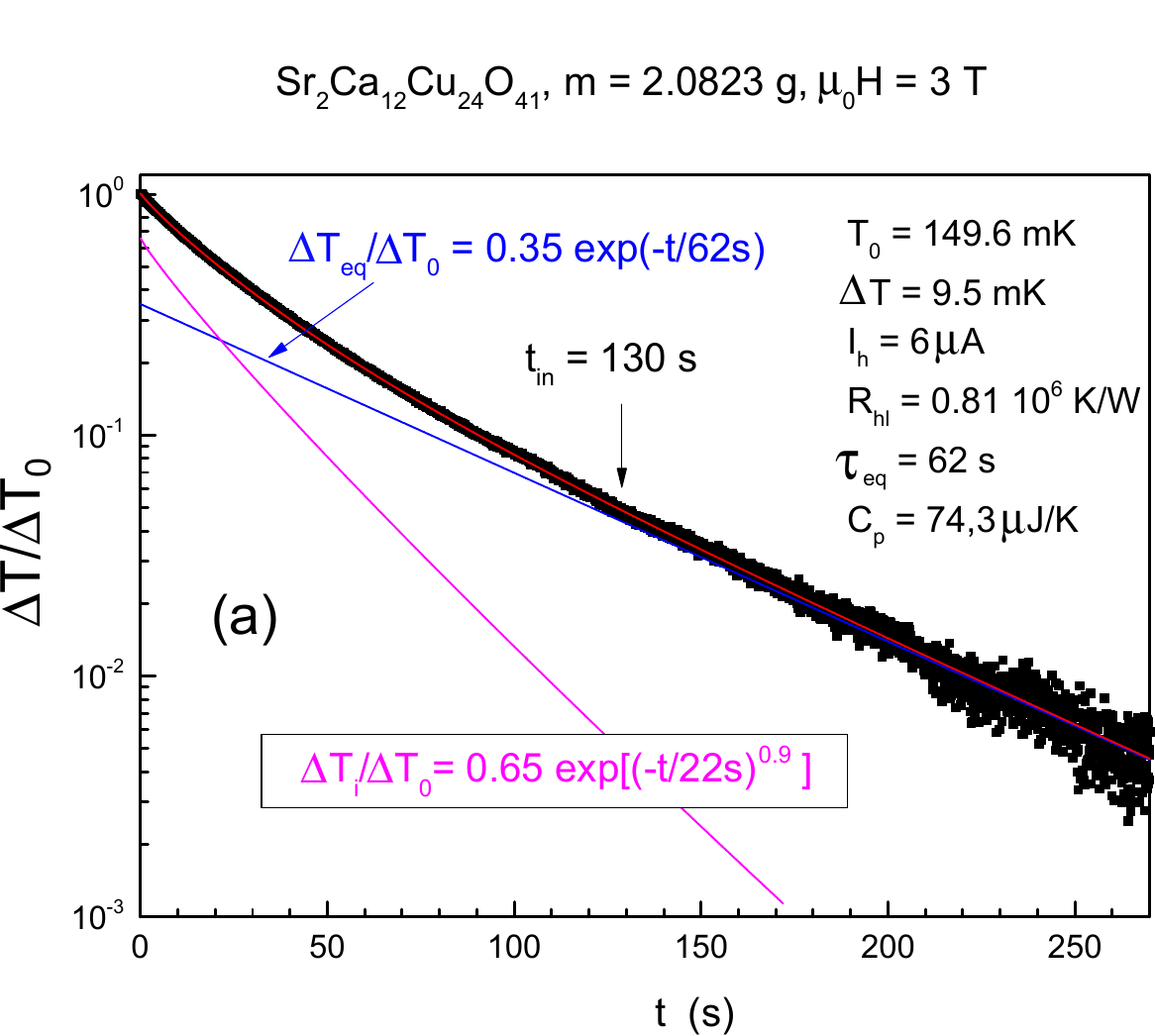}
	\includegraphics[width=7 cm]{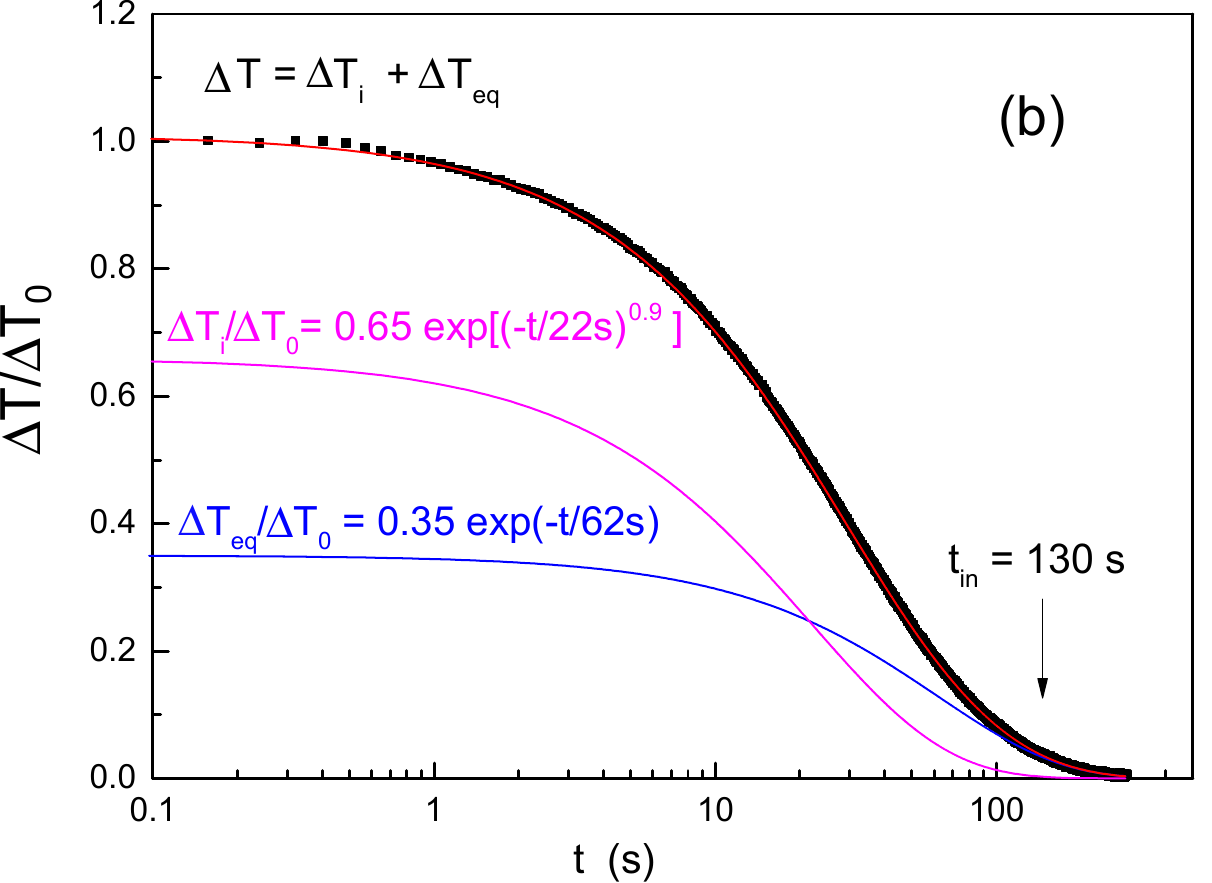}
	\includegraphics[width=7 cm]{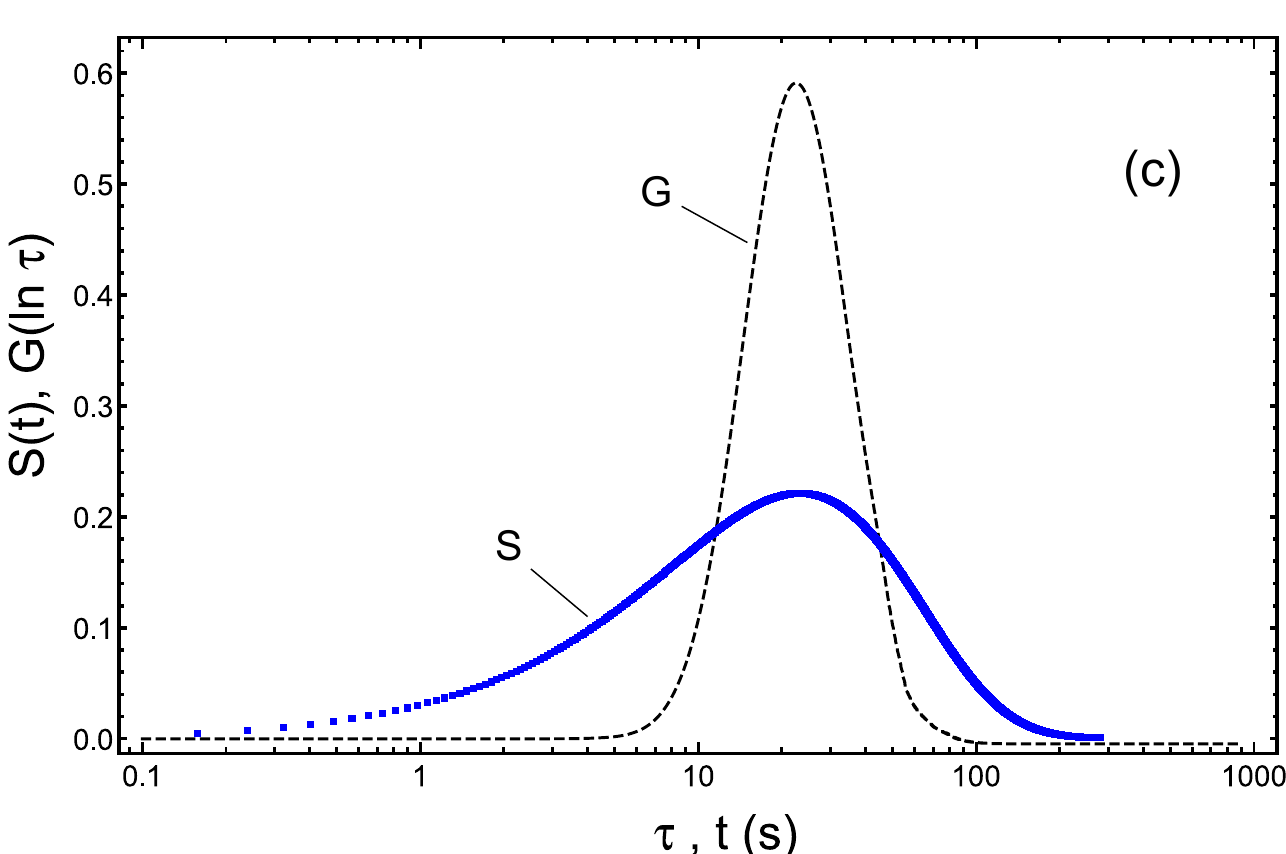}
	\caption{A  constant power of the sample heater is switched off at  $t = 0$  after the waiting time  $t_w = 2 t_{in}$  and the temperature relaxes from $160.1$ mK to $149.6$ mK. Figs.~11a  and 11b show the measured thermal relaxation $\Delta T(t)/\Delta T_0$ as a function of time in different semi-log presentations. The inner relaxation process is finished at $t_{in} = 130$ s. The thermal relaxation follows Eq.~(\ref{eq4}) at longer time $t > t_{in}$ with the quasi-equilibrium relaxation time $\tau_{eq} = 62$ s (see Fig.~11a). An additional thermal relaxation term $\Delta T_i/\Delta T_{i0}$ is observed at $t < t_{in}$ due to some inner relaxation process. We get a good fit of this term by a stretch exponential law with an exponent $\beta = 0.9$. This demonstrates that the width of the distribution function of relaxation time is small since  $\beta = 1$ corresponds to the case of a Delta-function  $\delta(\tau_m)$. The measured inner thermal relaxation $\Delta T_i/\Delta T_{i0}$ is shown in Fig. 11c (blue points) in the presentation $d(\Delta T_i/\Delta T_{i0})/d(\ln t)$ as a function of time for the data shown in Fig.~11 a,b. This gives roughly the distribution function of relaxation time (see the text). The numerical calculation of $\Delta T_i/\Delta T_{i0}$ with Eq.~(\ref{eq2}) yields a Gaussian distribution (dashed curve) with nearly the same position of the maximum $\tau_m$.  }
\end{figure}
 
\begin{table}\label{tb2}
\caption{Parameters of the temperature relaxation in Sr$_2$Ca$_{12}$Cu$_{24}$O$_{41}$ that are given by the expression $\Delta T_i/\Delta T_0 = A_i \exp[-(t/\tau_{mi})^{0.9}]$, $T_0 = 0.1$ K, $m = 2.0823$ g. Estimated errors for the given
	values are $10$~\%.}
\begin{tabular}{|c||r|r|r|r|}
\hline
$\mu_0 H$ & ~~~0 T~~~ & ~~~3 T~~~ & ~~~6 T~~~ & ~~~10 T~~~ \\
\hline
\hline
$A_i$ & 0.81 &0.81 & 0.81 & 0.81 \\
\hline
$\tau_{mi}~[s]$ & 63 & 143 & 252 & 543 \\
\hline
\end{tabular}
\end{table}

\begin{figure}[t]
	\includegraphics[width=8cm]{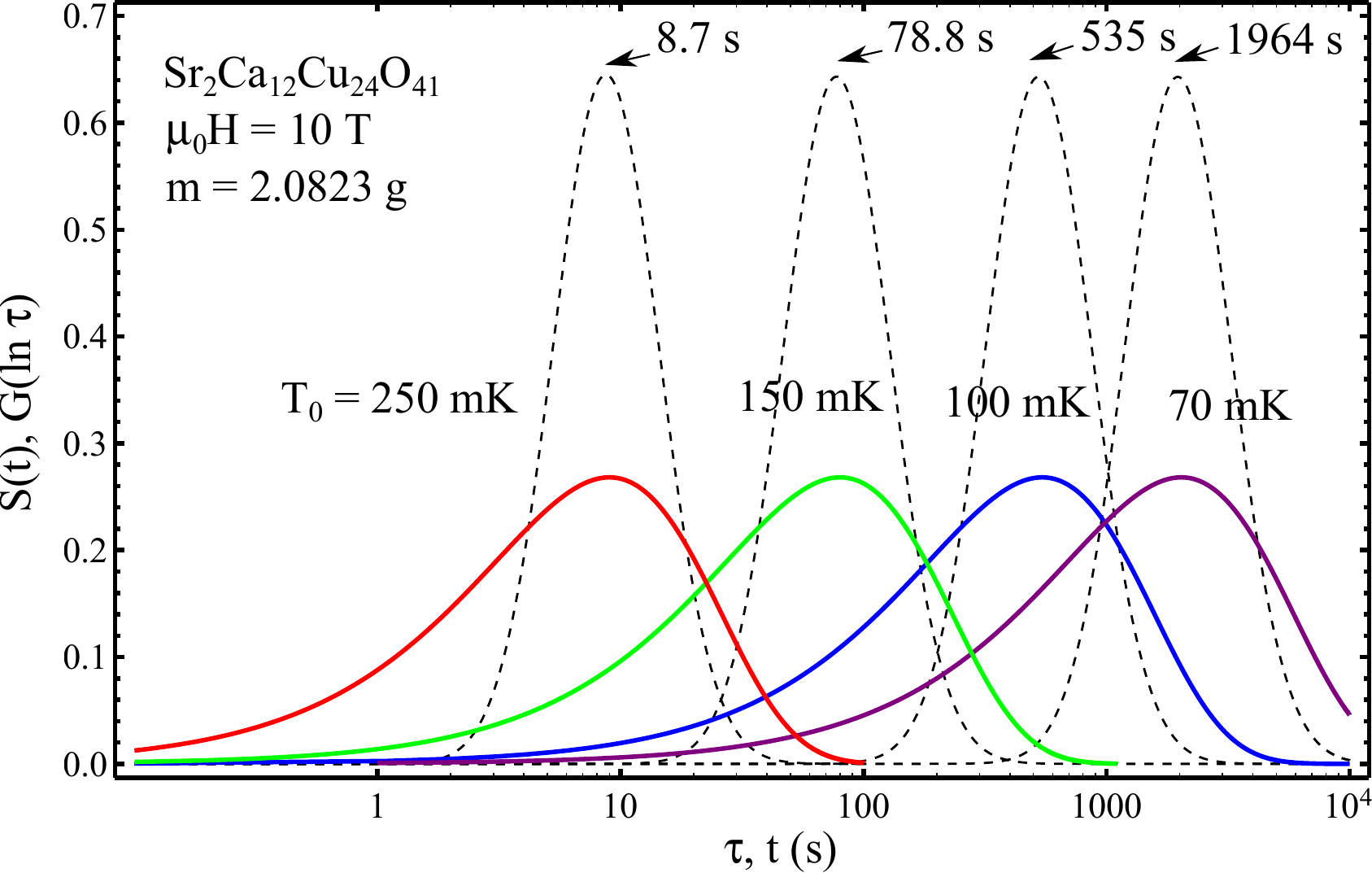}
	\caption{
	The measured inner thermal relaxation $\Delta T_i/\Delta T_{i0}$ in the presentation $d(\Delta T_i/\Delta T_{i0})/d\ln t$ as a function of time at $10$ T and different temperatures. The dashed curves show the distribution function obtained by the calculation of $\Delta T_i$ with Eq.~(\ref{eq2}). The width of the distribution function is independent from the magnetic field and temperature ($w = 0.5$). The temperature and magnetic field dependence of $\tau_m$ (the position of the maximum) is given in Fig.~13.}
\end{figure}

\section{Discussion}

The heat capacity is found to be very similar for all investigated materials with CDW or SDW: it is proportional to $T^{-2}$ and strongly time dependent.  However, the behavior of the relaxation time spectrum is very different for different materials: the distribution function contains one or two Gaussians, the width of the distribution is either a constant or dependent on the magnetic field, the relaxation time is usually caused by a thermal activation process, but in some materials tunneling seems to be possible at very low temperatures. 
\begin{figure}[h]
	\includegraphics[width=8cm]{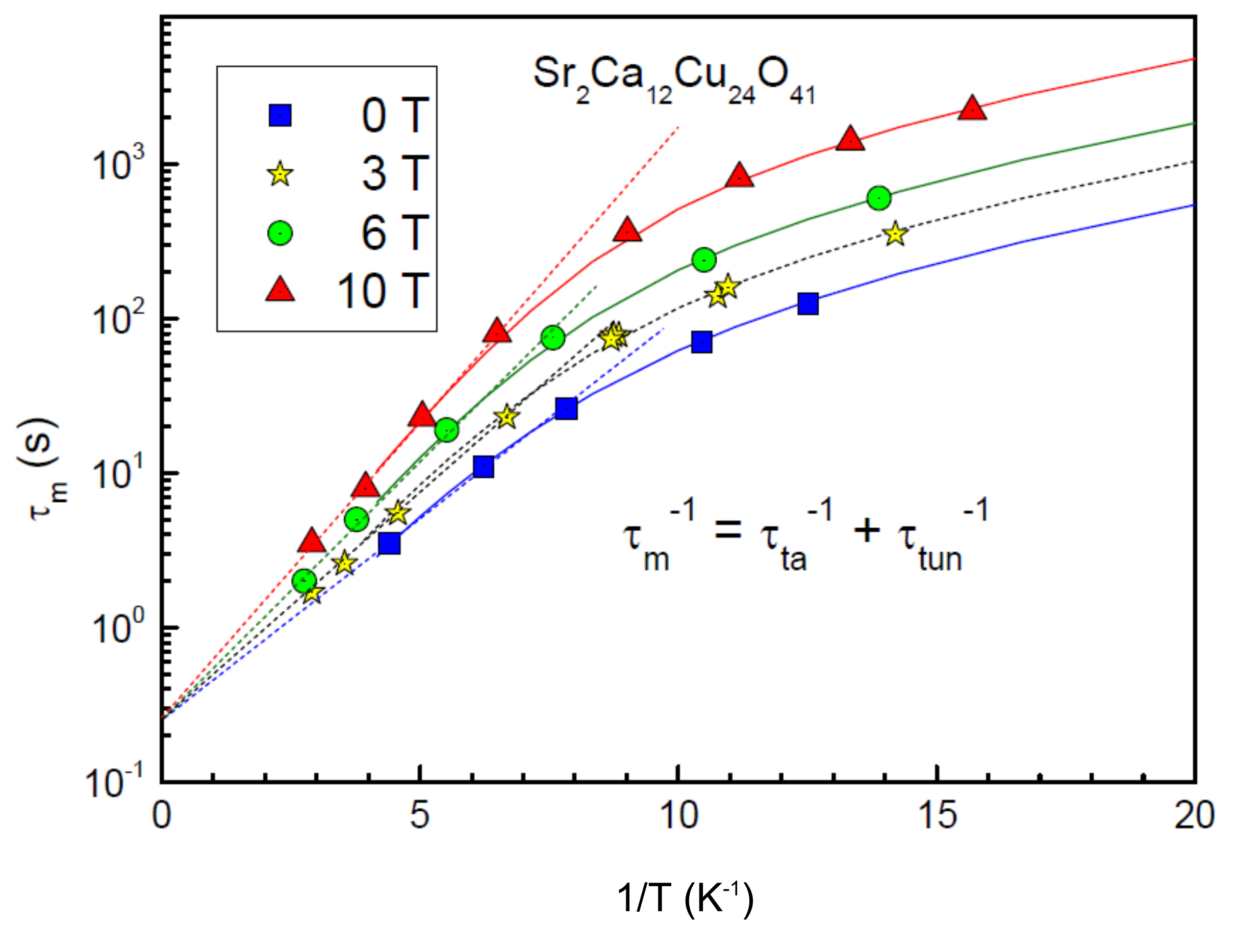}
	\caption{The Arrhenius plot of the peak position in the distribution function for different magnetic field. The activation energy increases with increasing magnetic field. Tunneling is observed at low temperatures. The curves are calculated for a relaxation rate determined by tunneling and thermal activated processes.}
\end{figure}

The analysis of the relaxation time spectra allows us to determine the distribution functions for all three materials. They demonstrate a rather specific behavior which gives an important information on the low-temperature ground state and collective effects. 

First of all, the experimental data for (TMTTF)$_2$Br confirm the previous results in Ref.~\cite{7c} and give a better understanding of the magnetic-field-induced density-wave glass state observed in Ref.~\cite{6c}. In particular, the distribution function is found to have a broad shape with low barrier height. At the same time, no tunneling process was observed. This means that there is a number of metastable states but the tunneling mass turns out to be large. The increasing magnetic field markedly shifts the position of the maximum (quadratic in H) while the width of the distribution function remains almost the same. What is important, the activation energy does not depend on the magnetic field.           

The long-time thermal relaxation in Sr$_{14}$Cu$_{24}$O$_{41}$ shows different behavior. The distribution function includes two well-defined narrow Gaussian peaks of different height. This indicates the presence of two types of relaxing structures. We observed a distinct difference in their behavior. The higher barriers shows both thermal activation and tunneling processes. They are weakly sensitive to the magnetic field. For lower barriers the relaxation time is caused by a thermal activation process only. The activation energy does not depend on $H$. As was mentioned in the introduction, this material develops the CDW ground state at low temperatures. Notice that according to the model of Larkin \cite{13c} and Ovchinnikov et al. \cite{14c} the low energy excitation appears as a consequence of pinning the CDW (or SDW) to defects. The strong impurity pinning leads to a generation of bisolitons, pairs of soliton and antisoliton \cite{6c}. Our study shows that there are two kinds of the effective two-level systems having a different origin. The nature of these states is still an open question.  

One of the most interesting results is the drastic change of the relaxation time spectrum in Sr$_{14}$Cu$_{24}$O$_{41}$ after doping the material with Ca: the maximum changes by a factor of $600$ (see Fig.~14). The distribution is represented by a single narrow Gaussian function. The relaxation includes both thermal and tunneling processes. A strong dependence on the magnetic field is found: we observed the exponential growth of the relaxation time due to a linear increase of the activation energy with H. Trying to understand such behavior, which differs markedly from both previous materials, let us recall that there is a principal difference between the low-temperature ground state of  Sr$_{14-x}$Ca$_{x}$Cu$_{24}$O$_{41}$ with $x=0$ and $x=12$. For $x=0$, the low temperature phase of the chains is 2D AF dimer and complementary charge order while for ladders it is the gapped spin-liquid and 2D CDW order. In case of $x=12$, the AF order is established both for the chains and the ladders~\cite{8c}. Therefore, the possible low energy excitations of Sr$_{2}$Ca$_{12}$Cu$_{24}$O$_{41}$ can be implemented through AF SDW transition followed by the Larkin-Ovchinnikov scenario of pinned SDW with the formation of bisolitons. The magnetic origin of bisolitons could explain the observed noticeable reaction to the magnetic field. This observation agrees with the findings in~\cite{18c} where the low-temperature magnetic response of these compounds is expected to be determined by the weakly coupled CuO$_2$ spin chains. A reduction of the number of holes in the chains through Ca doping leads to an additional contribution to the magnetization, which depends linearly on the magnetic field. What is important, the magnetization is found to be almost isotropic. In our experiments we directed H either along or normal to the c-axis. 

Notice that the narrow peak assumes the well-defined effective two-level systems, which supports the bisoliton picture with exactly two metastable states. This differs from the situation in (TMTTF)$_2$Br where presumably several metastable states appears due to local deformations of the SDW.  
The higher barriers in comparison to both (TMTTF)$_2$Br and Sr$_{14}$Cu$_{24}$O$_{41}$ can be explained as a result of a structural evolution. Indeed, it was found in~\cite{19c} that Ca doping increases the interaction between the two chain and ladder sublattices. According to the theoretical models in~\cite{13c,14c}, the barrier height growths with the interaction energy between chains.

\begin{figure}[t]
	\includegraphics[width=8cm]{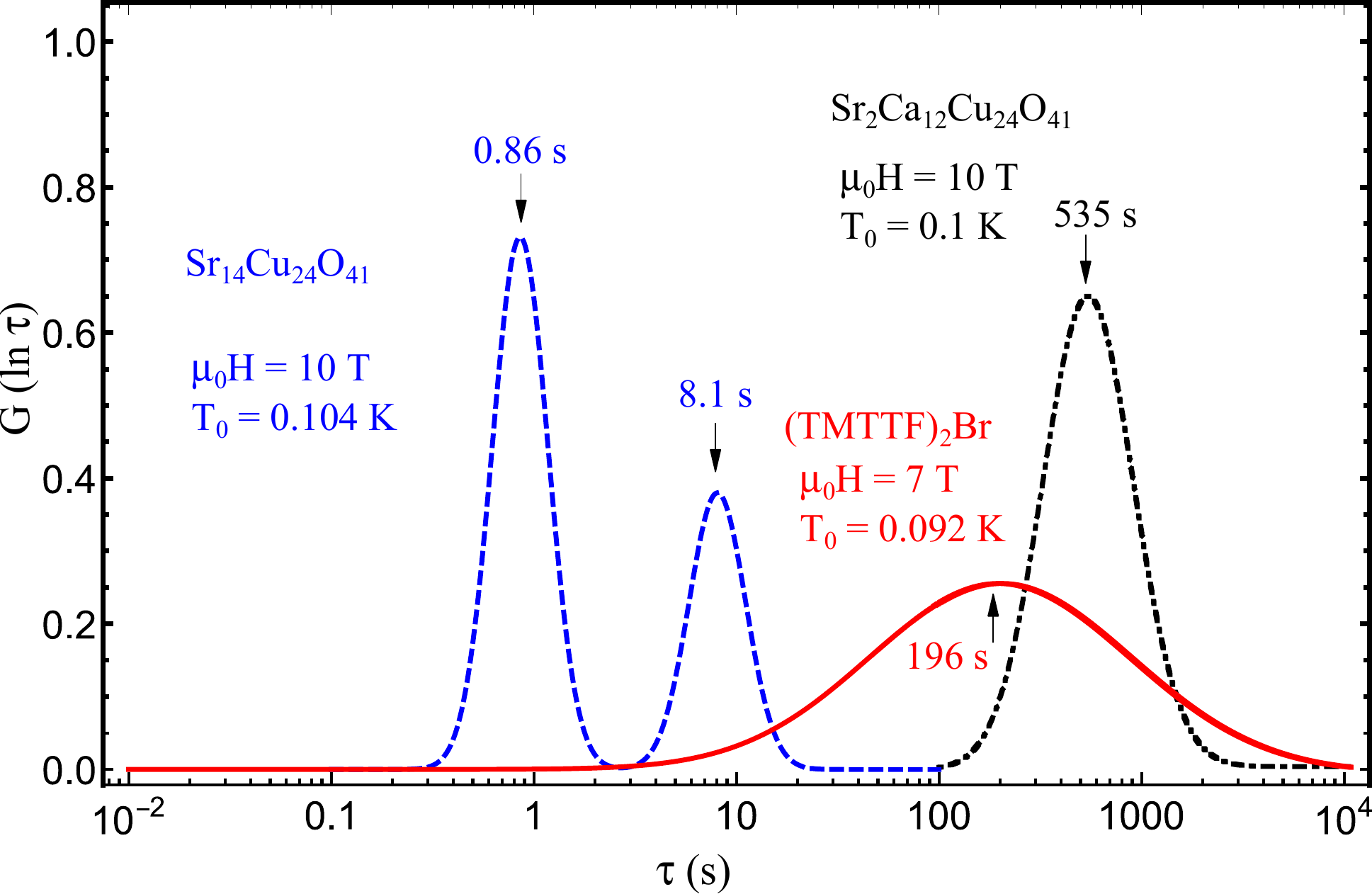}
	\caption{
		The distribution function of the relaxation time obtained for the three different materials at nearly the same temperature and high magnetic field. The doping of  Sr$_{14}$Cu$_{24}$O$_{41}$ with Ca shifts the distribution function to much longer time.}
\end{figure}

For comparison, the most important physical parameters  obtained in our experiments are given in Table III together with the results of Ref.~\cite{2c}. 
{\renewcommand{\arraystretch}{1.4}
\begin{table*}\label{tb3}
\caption{ \textbf{Most important parameters}\\}

\begin{threeparttable}
\begin{tabular}{|l||c|c|c|c|}
\hline
$~~~$ & (TMTTF)$_2$PF$_6$ & (TMTTF)$_2$Br & Sr$_{14}$Cu$_{24}$O$_{41}$& Sr$_2$Ca$_{12}$Cu$_{24}$0$_{41}$ \\
\hline
\hline
CDW/SDW & SDW & SDW & CDW & SDW \\
\hline
$E_a/ k_b$ [K] & 1.7 Ref.~\cite{2c} & 0.51\tnote{a} & 0.2 & 0.8-1.0\tnote{*} \\
\hline
$\tau_0 (0~\textrm{T})$ [s] & 0.42\tnote{**} & 0.05\tnote{a} & 0.2 and 0.5 & 0.3 \\
\hline
$\tau_0 (10~\textrm{T})$ [s] & - & 1.55 & 0.2 and 0.5 & 0.3 \\
\hline
$\tau_0(\mu_0 H > 4~\textrm{T})$  & -  & $\propto H$ & const. & const. \\
\hline
$\tau_m (0.1 \textrm{K}, 0 \textrm{T})$ [s] & 10$^7$\tnote{**} & 7.8 & 0.86 and 8.1 & 63 \\
\hline
$\tau_m (0.1 \textrm{K}, 10 \textrm{T})$ [s] & - & 242\tnote{b} & 0.86 and 8.1 & 535 \\
\hline
$E_s (4 \textrm{T})/ k_B$ [mK] & - & $<$ 52 & 40 & $<$ 35  \\
\hline
$E_s(\mu_0 H> 4 \textrm{T})$ [K] & - & $\propto H$ & $\propto H^2$ & $\propto H$ \\
\hline
$A_s $ [J/ mol K] & - & $>$ 50.5 & 3.9 & $>$ 3.9 \\
\hline
\end{tabular}
\begin{tablenotes}
       \item[*] magnetic field dependent
       \item[**] calculated from the maximum of the relaxation time spectrum at 0.2 K in Ref.~\cite{2c}.
       \item[a] for field induced phase
       \item[b] estimated with Eq.~(\ref{x2})
           \end{tablenotes}
  \end{threeparttable}
\end{table*}
Notice a remarkable change of the activation energy $E_a$ while its magnetic field dependence was observed in Sr$_2$Ca$_{12}$Cu$_{24}$0$_{41}$ only. 
The value of $\tau_0$ has the same order at zero field. For (TMTTF)$_2$Br it changes from $1$ s to $2.7$ s due to a phase transition and strong magnetic field dependence was observed. In other materials, no field dependence was found. $\tau_m$ does not depend on the magnetic field in Sr$_{14}$Cu$_{24}$O$_{41}$ and is found to be strongly dependent in two other materials. It should be noted that the origin of this dependence is different: it changes through $\tau_0$ in (TMTTF)$_2$Br and due to $E_a$ in Sr$_{14}$Cu$_{24}$O$_{41}$. $E_s$ is small in all investigated materials. Notice that Sr$_{14}$Cu$_{24}$O$_{41}$ is the first CDW/SDW system where the maximum of the Schottky term was observed. It is necessary to determine the absolute value of this contribution. Seemingly, there is a different magnetic field dependence for CDW  (as  $H^2$) and SDW (linear in $H$). Unfortunately, there is no chance to find the maximum of the heat capacity for incommensurate SDW systems since even at $10$ T it lies below 100 mK where $\tau_m$ exceeds 10$^7$ s! A similar situation takes place in the spin ice Dy$_2$Ti$_2$O$_7$: below $0.4$ K an additional time-dependent contribution to the heat capacity was observed, which could be the upper tail of a Schottky contribution. The inner thermal relaxation follows the Arrhenius law with a much larger activation energy  $E_a/k_B = 9.8$ K (see Ref.~\cite{Pomar}). The relaxation time is around $60000$ s at $0.34$ K and becomes too large below $0.3$ K to determine experimentally the equilibrium heat capacity. The amplitude $A_s$ in Sr$_{14}$Cu$_{24}$O$_{41}$ and Sr$_2$Ca$_{12}$Cu$_{24}$0$_{41}$ (roughly one defect per 2 unit cells) is much smaller than in (TMTTF)$_2$Br (6 defects per unit cell). A similar value was estimated from heat release data in (TMTTF)$_2$PF$_6$ (one defect per 4 unit cells, see Refs. \cite{5c,7c}). This large number is an indication that not impurities are the origin of two-level systems (TLSs).

Finally, let us compare the obtained distribution functions to those in other materials.  
For example, in structural glasses there was obtained a broad distribution in energy from $0$ to $k_B T$  and a broad distribution in $\tau$: $\Delta(\ln\tau)=\ln(\tau_{max}/\tau_{min})$ is found to have a value of about 50 for vitreous silica \cite{20c}. In CDW/SDW systems the distribution in $E$ is a Delta-function with a very low $E_s$. This indicates the presence of well-defined TLSs without any distribution. The distribution in $\tau $ is very small in comparison to a glass ($\Delta(\ln\tau)= 3$ for (TMTTF)$_2$Br and is equal to 1 for the other two investigated systems). In addition, there is a precise barrier height with some broadening. From this point of view, these materials are not glasses.

\section{Conclusion}

The most important result of the paper is the determination, for the first time, of the  distribution functions of relaxation times in systems with charge or spin density wave. As we have shown, this became possible by investigation of the thermal relaxation from a thermal equilibrium state at $T_1$ to another one at $T_0$ provided that the maximum time for inner relaxation $t_{in}$ is not too long. It has been found that a commonly used description based on the relaxation rate does not give the proper distribution function even in the case of a broad distribution (see TMTTF at short time). For the correct determination of the relaxation time distribution we have performed numerical calculations. 

We found the explicit form of distribution functions for three different materials 
with the result that they correspond to one or two Gaussian functions. A comparison of the temperature and magnetic field dependence of the relaxation time spectrum shows that the behavior is very different for studied materials. For (TMTTF)$_2$Br no tunneling process was observed while in two other materials both thermal and tunneling processes were revealed. We found a different reaction to the magnetic field: while the Schottky energy is found to have the same order of magnitude for all three materials (below 100 mK even at high magnetic field 10 T), it behaves as H$^2$ in TMTTF and Sr, but is linear in H for Ca. It should be also emphasized that 
the amplitude of the Schottky term, which is expected to be proportional to the number of pinning centers of the SDW or CDW~\cite{14c} turns out to be too large in (TMTTF)$_2$Br and more than 10 times smaller in Sr$_{14}$Cu$_{24}$O$_{41}$ and Sr$_{2}$Ca$_{12}$Cu$_{24}$O$_{41}$. This leads to the conclusion that low temperature anomalies in these materials are not universal. 

The most interesting experimental results concern Sr$_{2}$Ca$_{12}$Cu$_{24}$O$_{41}$. The fact is that the resistivity measurements reveal neither CDW nor SDW contribution at high Ca concentrations (see, e.g., ~\cite{8c}). Our experiments clearly show the relaxation process which includes both thermal activation and tunneling thus favoring the AF SDW transition in Sr$_{2}$Ca$_{12}$Cu$_{24}$O$_{41}$ at low temperatures. Notice that the resistivity measurements were performed at higher temperature and our method is much more sensitive. We expect that the more precise resistivity measurements must reveal the SDW contribution. 

The explicit form of the distribution function gives us an important information about the structure of the ground state and low-energy excitations in these materials. The theoretical model with two metastable states (SDW and bisolitons) gives reasonable explanation of our results for Sr$_{2}$Ca$_{12}$Cu$_{24}$O$_{41}$. In case of CDW compound Sr$_{14}$Cu$_{24}$O$_{41}$, there appears additional peak indicating on the presence of two kinds of the effective two-level systems having a different origin. The more complex situation takes place in (TMTTF)$_2$Br where the distribution function has a broad shape with low barrier height. There occurs several metastable states in SDW ground state. This case is beyond the theoretical description in~\cite{13c,14c} and needs for additional analysis. Perhaps this explains why the estimated number of pinning centers of the SDW is too large in this material.     

As the main conclusion, the investigation of the long time thermal relaxation yields more detailed information about the relaxation time spectrum, which lead after more systematical investigations to a better understanding of the low energy excitations in systems with CDW or SDW.

\textbf{Acknowledgement}
We acknowledge the support of the European Community Research Infrastructures under FP7 Capacities Specific Program, MICROCELVIN project number 228464 and the Landau-Heisenberg Program number HLP-2016-26.


\begin{thebibliography}{99}
	\bibitem{1c} J.C. Lasjaunias, K. Biljakovic,P. Monceau, Phys. Rev. B \textbf{53}, 7699 (1996).
	\bibitem{2c} J.C. Lasjaunias, R. Melin, D. Staresinic, K. Biljakovic and J. Souletie, Phys. Rev. Lett.
	{\textbf{94}}, 245701 (2005).
	\bibitem{3c} J.C. Lasjaunias, K. Biljakovic, S. Sahling, P. Monceau J. Phys. IV France {\textbf{131}}, 193 (2005).
	\bibitem{4c} J.C. Lasjaunias, S. Sahling, K. Biljakovic, P. Monceau, J. Magn. Mat.
	{\textbf{290-291}}, 989 (2005)
	\bibitem{5c}S. Sahling, J.C. Lasjaunias, K. Biljakovic, P. Monceau, J. Low Temp. Phys. {\textbf{133}}, 273 (2003).
	\bibitem{6c}R. Melin, J.C. Lasjaunias, S. Sahling, G. Remenyi and K. Biljakovic, Phys. Rev. Lett. {\textbf{97}}, 227203, (2006).
	\bibitem{7c} S. Sahling, J.C. Lasjaunias, R. Melin, P. Monceau and G. Remeyi. Eur, Phys. J. B. {\textbf{59}}, 9, (2007).
	\bibitem{Pomar} D. Pomaranski, L.R. Yaraskavitch, S. Meng, K.A. Ross, H.M.L. Noad,
	H.A. Dabkowska, B.D. Gaulin, and J.B. Kycia, Nat. Phys. {\textbf{9}}, 353, (2013).
		\bibitem{8c} T. Vuketic, B. Korin-Hamzic, T. Ivek, S. Tomic, B. Gorshunov, M. Dressel, J. Akimitsu, Physics  Reports \textbf{428}, 169 (2006).
	\bibitem{9c} S. Sahling et al., in work.
	\bibitem{10c} L. Lundgren, P. Svendlindh, P. Nordblad and O. Beckman, Phys. Rev. Lett. \textbf{51}, 911 (1983).
	\bibitem{11c} O. B\`{e}thoux, R. Brusetti, J.C. Lasjaunias and S. Sahling, Cryogenics \textbf{35}, 447 (1995).
	\bibitem{12c} S. Sahling, G. Remenyi, C. Paulsen, P. Monceau, V. Saligrama, C. Marin, A. Revcolevschi, L.P. Regnault, S. Raymond and. J.E. Lorenzo, Nature Physics \textbf{11}, 255 (2015).
	\bibitem{13c} A.I. Larkin, Zh. Eksp. Teor. Fiz. \textbf{105}, 1793 (1994).
	\bibitem{14c} Yu.N. Ovchinnikov, K. Biljakovic, J.C. Lasjaunias and. P. Monceau, EPL \textbf{34}, 649 (1996).
	\bibitem{15c} Y. Piskunov, D. Jerome, P. Auban-Senzier, P. Wzietek, A. Yakubovski, Phas. Rev. B \textbf{72}, 064512 (2005).
	\bibitem{16c} Y. Mizuno, T. Tohyama, S. Maekawa, J. Phys. Soc. Jpn. \textbf{66}, 937 (1997).
	\bibitem{17c} N. Nucker et al. Phys. Rev. B \textbf{62}, 14384 (2000).
	\bibitem{18c} R. Klingeler, N. Tristan, B. B\"{u}chner, M. H\"{u}cker, U. Ammerahl, and A. Revcolevschi, Phys. Rev. B \textbf{72}, 184406 (2005).
	\bibitem{19c} G. Deng, V. Pomjakushin, V. Petricek, E. Pomjakushina, M. Kenzelmann, and K. Conder, Phys. Rev. B \textbf{84}, 144111 (2011).
    \bibitem{20c} S. Sahling, S. Abens, and T. Eggert, Journal of Low Temperature Physics  \textbf{127}, 215 (2002)
\end{thebibliography}
\end{document}